\documentclass[preprint,nofootinbib]{revtex4}

\usepackage{amsmath}
\usepackage{graphicx}

\newcommand{\beq}{\begin{equation}}
\newcommand{\eeq}{\end{equation}}
\newcommand{\beqa}{\begin{eqnarray}}
\newcommand{\eeqa}{\end{eqnarray}}
\newcommand{\no}{\nonumber}
\def\OMIT#1{{}}
\newcommand{\lsim}{\mathrel{\rlap{\lower4pt\hbox{\hskip1pt$\sim$}}
    \raise1pt\hbox{$<$}}}         %less than or approx. symbol
\newcommand{\gsim}{\mathrel{\rlap{\lower4pt\hbox{\hskip1pt$\sim$}}
    \raise1pt\hbox{$>$}}}         %greater than or approx. symbol

\begin{document}

%\preprint{{\vbox{\hbox{}\hbox{}\hbox{}
%    \hbox{WIS/10/05-May-DPP}
%\hbox{hep-ph/0609178}}}}

\vspace*{.0cm}

\title{Probing CP violation in neutrino oscillations\\
with neutrino telescopes}

\author{Kfir Blum}\email{kfir.blum,yosef.nir,eli.waxman@weizmann.ac.il}
\affiliation{Department of Condensed Matter Physics,
  Weizmann Institute of Science, Rehovot 76100, Israel}
\author{Yosef Nir\footnote{The Amos de-Shalit chair of theoretical physics}}%\email{yosef.nir@weizmann.ac.il}
\affiliation{Department of Particle Physics,
  Weizmann Institute of Science, Rehovot 76100, Israel\vspace*{3mm}}
\author{Eli Waxman}%\email{eli.waxman@weizmann.ac.il}
\affiliation{Department of Condensed Matter Physics,
  Weizmann Institute of Science, Rehovot 76100, Israel\vspace*{3mm}}

%\date{\today}
%\pacs{}

\begin{abstract}%\vspace*{3mm}
Measurements of flavor ratios of astrophysical neutrino fluxes are
sensitive to the two yet unknown mixing parameters $\theta_{13}$ and
$\delta$ through the combination $\sin\theta_{13}\cos\delta$. We
extend previous studies by considering the possibility
that neutrino fluxes from more than a single type of sources
will be measured. We point
out that, if reactor experiments establish a lower bound on
$\theta_{13}$, then neutrino telescopes might establish an upper bound
on $|\cos\delta|$ that is smaller than one, and by that prove that CP
is violated in neutrino oscillations. Such a measurement requires
several favorable ingredients to occur: (i) $\theta_{13}$ is not far
below the present upper bound; (ii) The uncertainties in $\theta_{12}$ and
$\theta_{23}$ are reduced by a factor of about two; (iii) Neutrino
fluxes from muon-damped sources are identified, and their flavor
ratios measured with accuracy of order 10\% or better. For the last
condition to be achieved with the planned km$^3$ detectors, the
neutrino flux should be close to the Waxman-Bahcall bound. It
motivates neutrino telescopes that are effectively about
$10$ times larger than IceCube for energies of ${\cal O}(100\ TeV)$,
even at the expense of a higher energy threshold.
\end{abstract}

\maketitle

%%%%%%%%%%%%%%%%%%%%%%%%%%%%
\section{Introduction}
\label{sec:introduction}
One of the main goals of future neutrino experiments
\cite{Gonzalez-Garcia:2007ib} is to observe CP violation in neutrino
oscillations. The significance of such a measurement goes beyond the
determination of a fundamental parameter of Nature: it can give
further qualitative support to leptogenesis, the idea that the
observed baryon asymmetry of the Universe has its source in a lepton
asymmetry generated in neutrino interactions. In some scenarios, it is
even quantitatively related to leptogenesis.

Neutrino telescopes \cite{Halzen:2006mq}, such as the IceCube
experiment, aim to observe neutrinos coming from astrophysical
sources. The experiments will provide information on the direction,
energy, and flavor of the incoming neutrinos. In particular, ratios
between fluxes of different flavors arriving to the detector can be
measured. Ratios between these fluxes at the source are predicted by
rather robust theoretical considerations.

The modifications of the flavor ratios between source and detector
originate from neutrino oscillations. This means that the relations
between the fluxes at the source and the fluxes at the detector depend
on the neutrino parameters in a calculable way. Flavor measurements in
neutrino telescopes can thus provide information on the neutrino
mixing parameters
\cite{Farzan:2002ct,Pakvasa:2004hu,Serpico:2005sz,Winter:2006ce,Bhattacharjee:2005nh,Balaji:2006wi,Awasthi:2007az}.
In particular, there is sensitivity to two yet
unknown parameters: the mixing angle $\theta_{13}$ and the CP
violating phase $\delta$.

CP violation in neutrino oscillations can, in principle, be observed
via interference terms. For neutrinos coming from astrophysical
sources, such interference terms are washed out, and the measured
fluxes are therefore sensitive only to CP conserving parameters.
Specifically, the measured flavor ratios are sensitive to the
combination \beq\label{defdel}
\Delta_{13}\equiv\sin\theta_{13}\cos\delta. \eeq Since $\theta_{13}$
is experimentally bounded from above and known to be small, it is
convenient to write the flavor ratios in the general form
$a+b\Delta_{13}$, where $a$ and $b$ are known functions of the two
measured parameters, $\theta_{12}$ and $\theta_{23}$, but
independent of $\theta_{13}$ and $\delta$. The $b\Delta_{13}$ term
provides a small correction to the zeroth order prediction $a$. If
$\sin\theta_{13}=0$, or if CP violation is maximal, {\it i.e.}
$\delta=\pi/2$ or $3\pi/2$, the correction term is absent.

If $\sin\theta_{13}$ is close to the present experimental upper bound,
it is likely to be measured in near future reactor experiments
\cite{Ardellier:2004ui}. In that case, if neutrino telescopes are able
to exclude a correction term as large as $\pm b\sin\theta_{13}$, they
will establish that $\cos\delta\neq\pm1$ and, by that, will discover
that CP is violated in neutrino interactions.

Our goal in this paper is to analyze whether such a discovery of CP
violation by neutrino telescopes is at all possible. More
concretely, we do the following. On the qualitative level, we find
what types of sources and what types of flavor ratios provide the
strongest sensitivity to the parameters of interest. On the
quantitative level, we estimate the accuracy that is required in
these measurements and in independent measurements of the mixing
angles in order to establish that the CP violating phase is
different from $0$ and from $\pi$. Our final conclusion is that,
with large $\theta_{13}$ and near-maximal CP violation, and under
some favorable circumstances, it may be possible for IceCube (or,
more easily, for future, larger detectors) to establish CP violation
in neutrino interactions.

%%%%%%%%%%%%%%%%%
\section{Flavor ratios and mixing parameters}
\label{sec:anal}
Our goal in this section is to derive analytical expressions for
neutrino flavor fluxes that can be measured in neutrino
telescopes and, in particular, in IceCube.

Neutrino telescopes can identify the neutrino flavor
($\alpha=e,\mu,\tau$) via its characteristic interaction topology
\cite{Beacom:2003nh,Anchordoqui:2005is}. IceCube has an energy
threshold $\sim100\ GeV$ for detecting muon tracks, and $\sim1\ TeV$
for detecting electron- and tau-related showers. Above an energy
threshold $\sim1\ PeV$, it is possible to distinguish between the
electron-related electromagnetic showers and the tau-related hadronic
showers. Finally, around $E\sim 6.3\ PeV$, the Glashow resonance may
allow the identification of $\bar\nu_e$ events
\cite{Anchordoqui:2004eb,Bhattacharjee:2005nh}.

We denote the flux of $\nu_\alpha+\bar\nu_\alpha$ measured at the
detector by $\phi_\alpha^d$; the flux of antineutrinos
$\bar\nu_\alpha$ is denoted by $\bar\phi_\alpha^d$.
We consider the following flavor ratios:
\beqa\label{defr}
R&\equiv&\frac{\phi_\mu^d}{\phi_e^d+\phi_\tau^d},\\
\label{defs}
S&\equiv&\frac{\phi_e^d}{\phi_\tau^d},\\
\label{deft}
T&\equiv&\frac{\bar\phi_e^d}{\phi_\mu^d}.
\eeqa
Below $E\sim PeV$, only $R$ can be measured. At higher energies, $S$
and perhaps $T$ may become available.

We denote the
flux of $\nu_\alpha+\bar\nu_\alpha$ emitted from the source by
$\phi_\alpha^s$. The relation between $\phi_\alpha^s$ and
$\phi_\beta^d$ is given by
\beq\label{soudet}
\phi_\beta^d = P_{\beta\alpha}\phi_\alpha^s,
\eeq
where $P_{\beta\alpha}\equiv P(\nu_\alpha\to\nu_\beta)$ is the
transition probability from a flavor $\nu_\alpha$ at the source to a
flavor $\nu_\beta$ at the detector.

For propagation over astronomical distance scales, the
distance-dependent oscillatory terms average out, and
$P_{\beta\alpha}$ depends on mixing parameters only:
\beq\label{pabu}
P_{\beta\alpha}=\sum_i|U_{\alpha i}|^2|U_{\beta_i}|^2.
\eeq
Here $U$ is the unitary transformation that relates the neutrino
interaction eigenstates $\nu_\alpha$ ($\alpha=e,\mu,\tau$) and mass
eigenstates $\nu_i$ ($i=1,2,3$):
\beq
|\nu_\alpha\rangle=U_{\alpha i}^*|\nu_i\rangle.
\eeq
We parametrize the matrix $U$ by three mixing angles, $\theta_{12}$,
$\theta_{23}$ and $\theta_{13}$, and three CP violating phases,
$\delta$, $\alpha_1$ and $\alpha_2$:
\beq
U=\begin{pmatrix} c_{12}c_{13}&s_{12}c_{13}&s_{13}e^{i\delta}\cr
  -s_{12}c_{23}-c_{12}s_{23}s_{13}e^{i\delta}&
  c_{12}c_{23}-s_{12}s_{23}s_{13}e^{i\delta}&
  s_{23}c_{13}\cr
  s_{12}s_{23}-c_{12}c_{23}s_{13}e^{i\delta}&
  -c_{12}s_{23}-s_{12}c_{23}s_{13}e^{i\delta}&
  c_{23}c_{13}\cr\end{pmatrix} \begin{pmatrix}
  e^{i\alpha_1/2}&&\cr &e^{i\alpha_2/2}&\cr &&1\cr\end{pmatrix},
\eeq
where $c_{ij}\equiv\cos\theta_{ij}$, $s_{ij}\equiv\sin\theta_{ij}$. It
is clear from Eq. (\ref{pabu}) that $P_{\beta\alpha}$ is independent
of the phases $\alpha_{1,2}$. It depends on the three mixing angles
$\theta_{ij}$ and on $\delta$.

Since it is experimentally known that $\theta_{13}$ is small (see
Table \ref{tab:angles}), it is convenient to write down the flavor
transition probabilities to first order in $\Delta_{13}$ (see
Eq. (\ref{defdel})) \cite{Xing:2006uk,Xing:2006xd,Rodejohann:2006qq}:
\beqa\label{trapro}
P_{ee}&\simeq&1-\frac12\sin^22\theta_{12},\no\\
P_{e\mu}&\simeq&\frac12\sin^22\theta_{12}\cos^2\theta_{23}
+\frac14\sin2\theta_{23}\sin4\theta_{12}\Delta_{13},\no\\
P_{\mu\mu}&\simeq&1-\frac12\left(
  \cos^4\theta_{23}\sin^22\theta_{12}+\sin^22\theta_{23}\right)
-\frac12\sin2\theta_{23}\cos^2\theta_{23}\sin4\theta_{12}\Delta_{13},\no\\
P_{e\tau}&\simeq&\frac12\sin^22\theta_{12}\sin^2\theta_{23}
-\frac14\sin2\theta_{23}\sin4\theta_{12}\Delta_{13},\no\\
P_{\mu\tau}&\simeq&\frac18\sin^22\theta_{23}\left(4-
  \sin^22\theta_{12}\right)
+\frac18\sin4\theta_{23}\sin4\theta_{12}\Delta_{13}.
\eeqa
The remaining probabilities can be derived from
$P_{\alpha\beta}=P_{\beta\alpha}$ and $\sum_\alpha
P_{\alpha\beta}=1$.

%%%%%%%%%%%%%%%%%%
\section{Astrophysical neutrino sources and flavor ratios}
We consider two types of sources:
\begin{itemize}
\item ``Pion sources'' (denoted by sub-index $\pi$) provide the following
  flavor ratios:
  \beq\label{phipi}
  \phi_e^s:\phi_\mu^s:\phi_\tau^s=1:2:0.
  \eeq
As concerns the $\nu_e-\bar\nu_e$ decomposition of $\phi_e$, the
situation depends on whether the pions are produced mainly by $pp$ or
$p\gamma$ interactions:
\beq\label{bphipi}
\frac{\bar\phi_\mu^s}{\phi_\mu^s}=\frac12,\ \ \ \ \
\frac{\bar\phi_e^s}{\phi_e^s}=\begin{cases}
  1/2&pp,\cr 0&p\gamma.\cr
\end{cases}
\eeq
\item ``Muon-damped sources'' (denoted by sub-index $\mu$) provide the
  following flavor ratios:
  \beq\label{phimu}
  \phi_e^s:\phi_\mu^s:\phi_\tau^s=0:1:0.
  \eeq
As concerns the $\nu_\mu-\bar\nu_\mu$ decomposition of $\phi_\mu$, the
situation depends on whether the pions are produced mainly by $pp$ or
$p\gamma$ interactions:
\beq\label{bphimu}
\frac{\bar\phi_\mu^s}{\phi_\mu^s}=\begin{cases}
  1/2&pp,\cr 0&p\gamma.\cr
\end{cases}
\eeq
\end{itemize}

The expectation is that all sources where the initial stage of
neutrino production is charged pion decays will undergo a transition
from a ``pion'' to ``muon-damped'' flavor decomposition at high enough
neutrino energies \cite{Kashti:2005qa}. If energy losses are mainly
due to synchrotron radiation and inverse compton emission, the
transition region is expected to span about one decade in energy. The
actual threshold energy cannot be determined model independently and,
furthermore, is likely to differ from source to source. We assume here
that, nevertheless, the transition is such that it will be possible to
separate the neutrino events to lower-energy events from pion sources
and higher-energy events from muon-damped sources.

The dependence of the flavor ratios at the detector on the mixing
parameters can be obtained as follows. One starts from the fluxes at
the source (in arbitrary units), Eqs. (\ref{phipi}), (\ref{bphipi}),
(\ref{phimu}) and (\ref{bphimu}). Then, the fluxes at the detector
can be found by using Eq. (\ref{soudet}) and the expressions for the
transition probabilities (\ref{pabu}). Finally, the expressions are
put in Eqs. (\ref{defr}), (\ref{defs}) and (\ref{deft}).

\section{Description of Analysis}
\label{sec:des}
%%%%%%%%%%%%%%%%%%
\subsection{Numerical input}
\label{sec:num}
The current best fit values and $1\sigma$ ranges of the mixing
angles are given in Table \ref{tab:angles}
\cite{Gonzalez-Garcia:2007ib}. By the time that IceCube can carry
out the measurements that we discuss in this work, it is likely that
the knowledge -- from other experiments -- of the mixing angles will
improve. Such progress is very significant for our purposes, as we
see below. In particular, in order that the IceCube measurements
will be able, even in principle, to show that $\delta\neq0$, it is
crucial that experiments establish that $\sin\theta_{13}\neq0$. For
the sake of our analysis, we assume that reactor experiments will
measure $\sin^22\theta_{13}=0.090\pm0.013$ \cite{Ardellier:2004ui}.
(This value for $\theta_{13}$ corresponds to the current $2\sigma$
allowed range \cite{Gonzalez-Garcia:2007ib}.) For $\theta_{12}$ and
$\theta_{23}$ we assume a factor of two improvement in the accuracy.
The resulting ranges which we use to examine the question of whether
IceCube can discover CP violation are given in the column labelled
`Future' in Table \ref{tab:angles}. As concerns the phase, we assume
that it will remain unconstrained.

%\begin{table}[t]  %%%% note - kfir 10/06/2007 - replace by 1sigma ranges.
%\caption{Experimental ranges of mixing angles
%  \cite{Gonzalez-Garcia:2007ib}}
%\label{tab:angles}
%\begin{center}
%\begin{tabular}{c|cc|c} \hline\hline
%\rule{0pt}{1.2em}%
%%\label{tab:bqqq}
%%\settabs 5 \columns
%Parameter & Best fit & $3\sigma$ range  & `Future' \cr \hline\hline
%$\sin^2\theta_{12}$ & $0.31$ & $0.25-0.38$ & $\ 0.31\pm0.03$  \cr
%$\sin^2\theta_{23}$ & $0.47$ & $0.32-0.64$ & $\ 0.47\pm0.08$  \cr
%$\sin^2\theta_{13}$ & $0.00$ & $\leq0.04$ & $\ 0.022\pm0.037$  \cr
%\hline\hline
%\end{tabular}
%\end{center}
%\end{table}
\begin{table}[t]
\caption{Experimental ranges of mixing angles
  \cite{Gonzalez-Garcia:2007ib}}
\label{tab:angles}
\begin{center}
\begin{tabular}{c|cc|c} \hline\hline
\rule{0pt}{1.2em}%
%\label{tab:bqqq}
%\settabs 5 \columns
Parameter & Best fit & $1\sigma$ range  & `Future'\cr \hline\hline
$\sin^2\theta_{12}$ & $0.31$ & $0.29-0.33$ & $\ 0.31\pm0.01$  \cr
$\sin^2\theta_{23}$ & $0.47$ & $0.40-0.55$ & $\ 0.47\pm0.04$  \cr
$\sin^2\theta_{13}$ & $0.00$ & $\leq0.008$ & $\ 0.022\pm0.003$  \cr
\hline\hline
\end{tabular}
\end{center}
\end{table}

To obtain an understanding of the dependence of the flavor ratios on
the mixing parameter $\Delta_{13}$, we use the central values for
the two measured angles, $\theta_{12}$ and $\theta_{23}$, and apply
the approximate relations (\ref{trapro}). We obtain for the pion
source
\beqa\label{piobs}
R_\pi&=&0.49-0.15\Delta_{13},\no\\
S_\pi&=&1.04+0.52\Delta_{13},\no\\
T_\pi&=&\begin{cases}
  0.52+0.28\Delta_{13}&pp,\cr
  0.23+0.22\Delta_{13}&p\gamma,\cr
  \end{cases}
\eeqa
and for the muon-damped source
\beqa\label{muobs}
R_\mu&=&0.62-0.49\Delta_{13},\no\\
S_\mu&=&0.58+0.44\Delta_{13},\no\\
T_\mu&=&\begin{cases}
  0.30+0.38\Delta_{13}&pp,\cr
  0&p\gamma,\cr
  \end{cases}
\eeqa

We emphasize, however, that in our calculations we use the full
dependence on the mixing angles [see Eq. (\ref{pabu})], and not just
the leading order (in $\Delta_{13}$) expressions, Eqs. (\ref{trapro}),
(\ref{piobs}) and (\ref{muobs}).

%%%%%%%%%%%%%%%%%%%%%%%%%%
\subsection{Experimental errors}
\label{sec:experr}
It is not yet clear whether all of the flavor ratios defined in
Section \ref{sec:anal} will indeed be available at IceCube (or any
future neutrino telescope). We assume that $R_\pi$, $R_\mu$ and
$S_\mu$ will be measured, and consider cases where $S_\pi$ and $T_\mu$
are available or not.

The goal of this work is not to obtain a detailed realistic estimate
of the accuracies that are {\it expected} in the relevant
measurements. Such an estimate depends on both features of the
astrophysical neutrinos that are not yet known ({\it e.g.} the actual
total flux), and features of the detectors that will only become clear
when these neutrinos are observed. The main goal here is to find the
accuracies that are {\it required} in order to establish that CP is
violated.

We thus consider the following experimental accuracies in the
measurements of the various flavor ratios:
\begin{enumerate}
\item $R_\pi$: we consider hypothetical accuracies of $5\%$, $10\%$ or
  $20\%$. If the flux is close to the Waxman-Bahcall bound, then we
  expect ${\cal O}(100)$ events, and an error of order $10\%$ seems
  realistic;
\item $S_\pi$: In the cases that it is available, we relate the
  accuracy to that of $R_\pi$, by assuming a Poisson distribution of
  the number of events for each neutrino flavor. We neglect issues of
  efficiency in detecting tracks versus showers. This leads to $\Delta
  S_\pi/S_\pi=\sqrt{S_\pi(1+S_\pi^{-1})^2/(1+R_\pi^{-1})}(\Delta
  R_\pi/R_\pi)$. Using central values from Eq. (\ref{piobs}), we
  obtain $\Delta S_\pi/S_\pi=1.2(\Delta R_\pi/R_\pi)$;
\item $R_\mu$: we consider hypothetical accuracies which are at best
  the same as the error on $R_\pi$ and at worst $20\%$;
\item $S_\mu$: Following the same line of thought as for $S_\pi$, we
  use $\Delta
  S_\mu/S_\mu=\sqrt{S_\mu(1+S_\mu^{-1})^2/(1+R_\mu^{-1})}(\Delta
  R_\mu/R_\mu)$. Using central values from Eq. (\ref{muobs}), we
  obtain $\Delta S_\mu/S_\mu=1.3(\Delta R_\mu/R_\mu)$;
\item $T_\mu$: In the cases that it is available, we assume $\Delta
  T_\mu/T_\mu=\Delta R_\mu/R_\mu$.
\end{enumerate}

The various scenarios can be defined by the assumed accuracies in
$R_\pi$ and $R_\mu$: We denote by $(a,b)$ a scenario where the errors
are $\Delta R_\pi/R_\pi=a\%$ and $\Delta R_\mu/R_\mu=b\%$. The six
scenarios that we consider are presented in Table \ref{tab:models}.

\begin{table}[t]
\caption{Scenarios for experimental accuracies}
\label{tab:models}
\begin{center}
\begin{tabular}{l|ccccc} \hline\hline
\rule{0pt}{1.2em}%
%\label{tab:bqqq}
%\settabs 5 \columns
Scenario & $\Delta R_\pi/R_\pi$ & $\Delta R_\mu/R_\mu$ & $\Delta
S_\mu/S_\mu$ & ($\Delta S_\pi/S_\pi$) & ($\Delta T_\mu/T_\mu$) \cr \hline\hline
$(5,5)$ & $5$ & $5$ & $7$ & $6$ & $5$  \cr
$(5,10)$ & $5$ & $10$ & $13$ & $6$ & $10$  \cr
$(5,20)$ & $5$ & $20$ & $27$ & $6$ & $20$  \cr
$(10,10)$ & $10$ & $10$ & $13$ & $12$ & $10$  \cr
$(10,20)$ & $10$ & $20$ & $27$ & $12$ & $20$  \cr
$(20,20)$ & $20$ & $20$ & $27$ & $24$ & $20$  \cr \hline\hline
\end{tabular}
\end{center}
\end{table}

We thus consider a hypothetical set of measurements -- $R$, $S$, $T$
and $\sin^2\theta_{ij}$ -- which provide information on
$\theta_{ij}$ and $\delta$. The statistical procedure by which this
information is extracted is described in the following section.

%%%%%%%%%%%%%%%%%%%%%%%%%%
\subsection{Statistical procedure}
\label{sec:stat}
Given a measurement of an observable $Y^{\rm meas}=\langle
Y\rangle\pm\sigma_Y$, we construct
$\chi^2(\theta_{ij},\delta)=\sum_Y\left[\frac{\langle
    Y\rangle-Y(\theta_{ij},\delta)}{\sigma_Y}\right]^2$, where $Y(\theta_{ij},\delta)$
represents the theoretical description of the $Y$ observable. The
uncertainty $\sigma_Y$ is given in Table \ref{tab:angles} for
$\sin^2\theta_{ij}$ and in Table \ref{tab:models} for $R$, $S$ and
$T$. A statistical handling of the parameters is performed by
analyzing the quantity
$\Delta\chi^2(\theta_{ij},\delta)=\chi^2-\min_{\theta_{ij},\delta}\{\chi^2\}$.

We define the $N$-dimensional ``$\alpha$\% CL acceptance region'', for
a subset of $N$ out of the four mixing parameters
($\theta_{ij},\delta$), by the region in the $N$ parameter space for
which $\Delta\chi^2_{\rm marg}<C^{-1}(\alpha,N)$. Here
$\Delta\chi^2_{\rm marg}$ is obtained by marginalizing $\Delta\chi^2$
with respect to the $4-N$ redundant parameters and $C^{-1}(\alpha,N)$
is the inverse chi-square CDF with $N$ degrees of freedom, evaluated
at the point $\alpha$. We have compared this procedure to the more
computationally demanding FC construction, (as described in
\cite{Feldman:1997qc} and demonstrated, for example, in
\cite{Schwetz:2006md}) under the assumption of gaussian measurement
errors, for several sample configurations. We have found a reasonable
agreement between our simplified method and the full FC routine, with
the former tending in general to supply slightly more conservative
acceptance regions.

We define the ``$\alpha$\% CL acceptance interval'', for a specific
parameter, by the set of parameter values for which the condition
$\Delta\chi^2_{\rm marg}<C^{-1}(\alpha,1)$ is satisfied, with
$\Delta\chi^2_{\rm marg}$ given by marginalizing $\Delta\chi^2$ with
respect to all of the other parameters.

An ``$\alpha$\% CL fraction of coverage'' is further defined for a
specific parameter as the percentage of the parameter range that is
included in the $\alpha$\% CL acceptance interval. The lower is this
fraction, the stronger is the exclusion power of the experiment with
respect to the relevant parameter.

We say that a specific value of a parameter is excluded with
$\alpha$\% confidence, if this value is not contained in the
corresponding $\alpha$\% acceptance interval. This notion will be
used below, when we discuss the prospects of various measurement
scenarios do exclude CP conservation in neutrino oscillations.

%%%%%%%%%%%%%%%%%%%%%%%%%%
\section{Results}
\label{sec:results}
\subsection{Neglecting uncertainties in $\theta_{12}$ and
  $\theta_{23}$}
\label{sec:neg}
To understand the abilities and difficulties that are intrinsic to the
measurements by neutrino telescopes, we first carry out an analysis
where $\theta_{12}$ and $\theta_{23}$ are held fixed at their
current best fit values. In the next section, we will study the
implications of the uncertainties in these angles.

We begin by choosing specific values for the parameters
$\theta_{13}$ and $\delta$, which we call ``true parameters''.
Concretely, we assume a true value $\theta_{13}=0.15$, and consider
mainly three possibilities for the true value of $\delta$: the two
CP conserving ones ($\delta=0,\pi$) and the maximally CP violating
one ($\delta=\pi/2$). We evaluate the flux ratios that theoretically
correspond to these mixing parameters. For the sake of illustration,
we assume that the experimental measurements will obtain these flux
ratios as their central values, with errors as specified for each of
our six scenarios. We then perform a fit to $\theta_{13}$ and
$\delta$ (obtaining, of course, the ``true values'' as the best-fit
parameters, but with acceptance regions that are different between
the various scenarios).

\begin{figure}[hbp]
\hspace{-2cm} $\begin{array}{lcr}
\includegraphics[width=80mm,height=60mm]{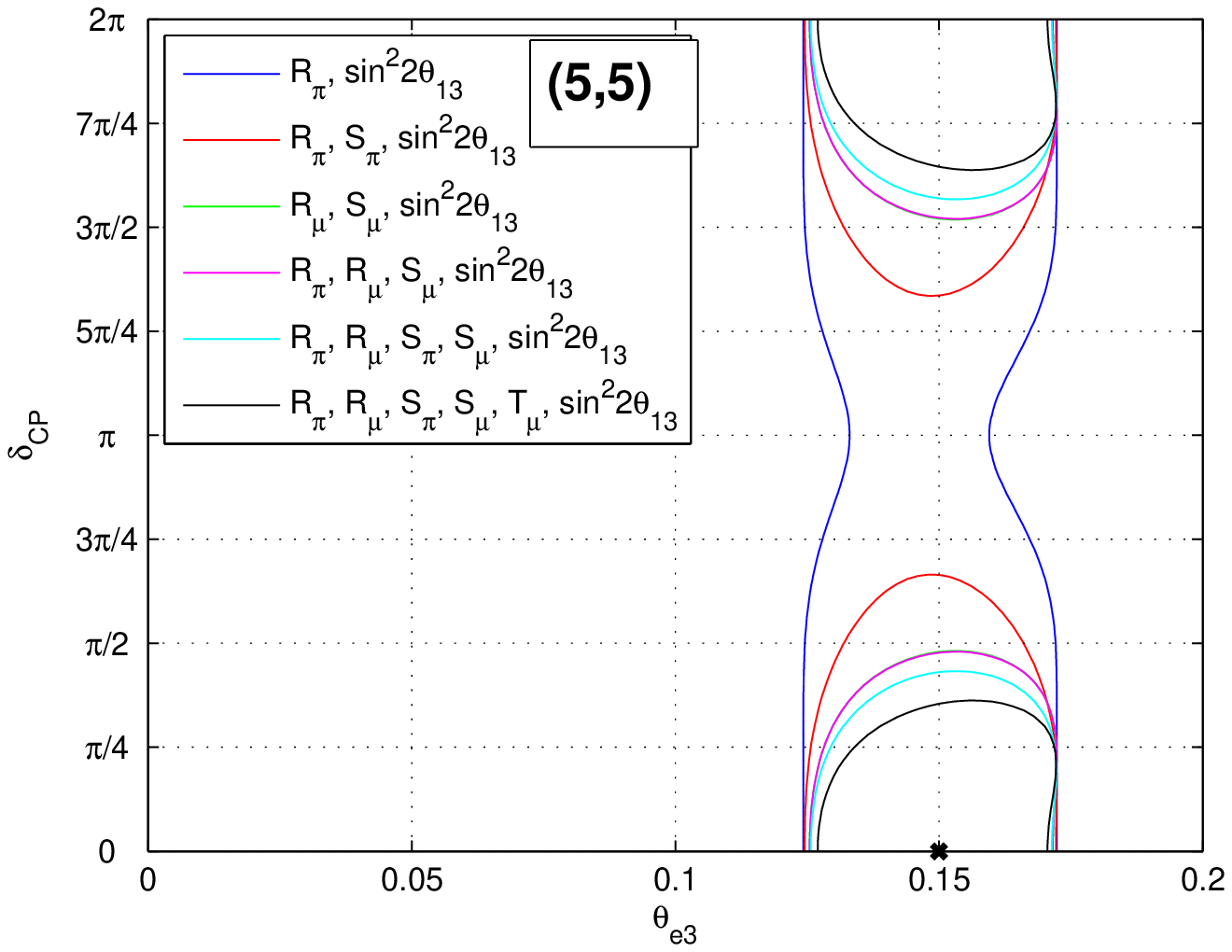}  &
\includegraphics[width=80mm,height=60mm]{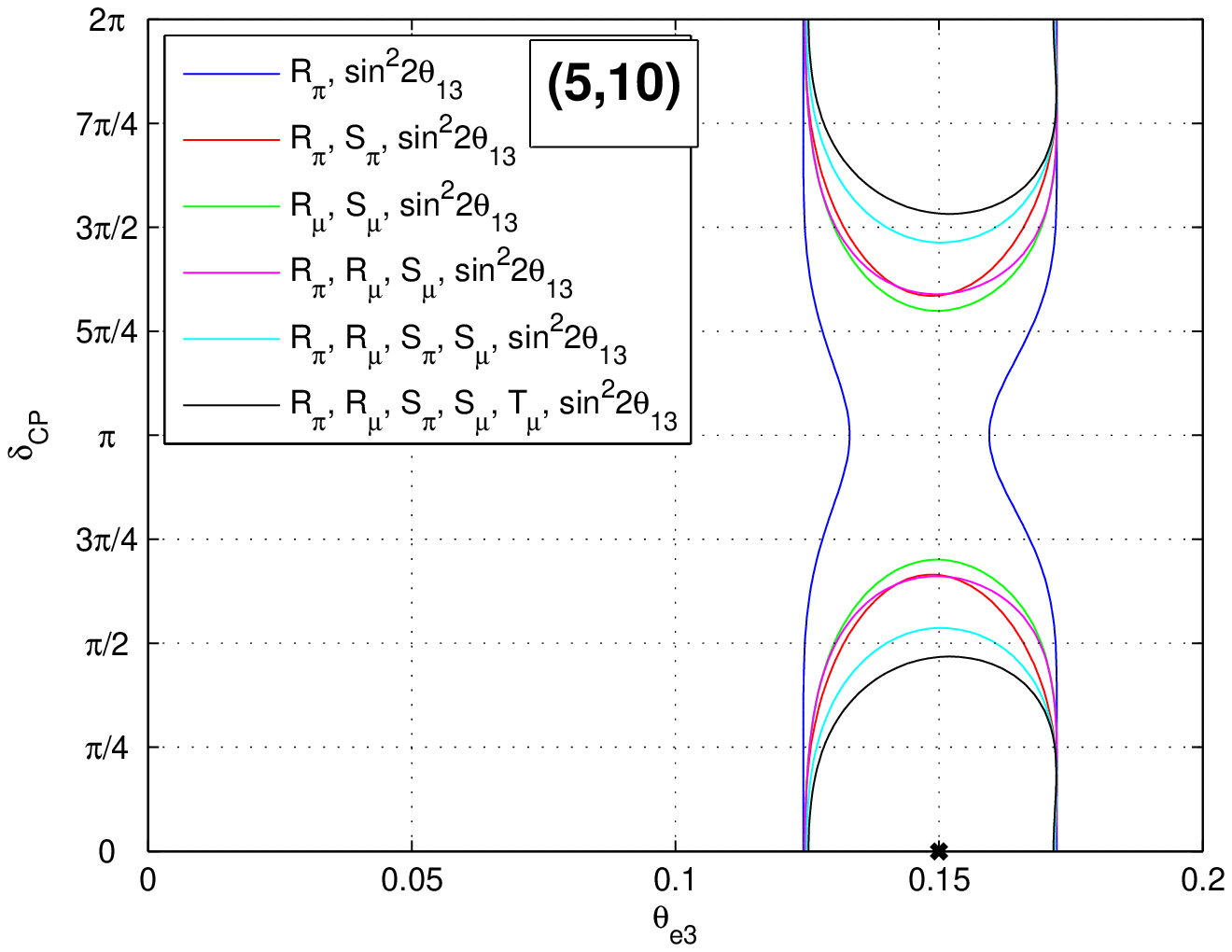}  \\
\includegraphics[width=80mm,height=60mm]{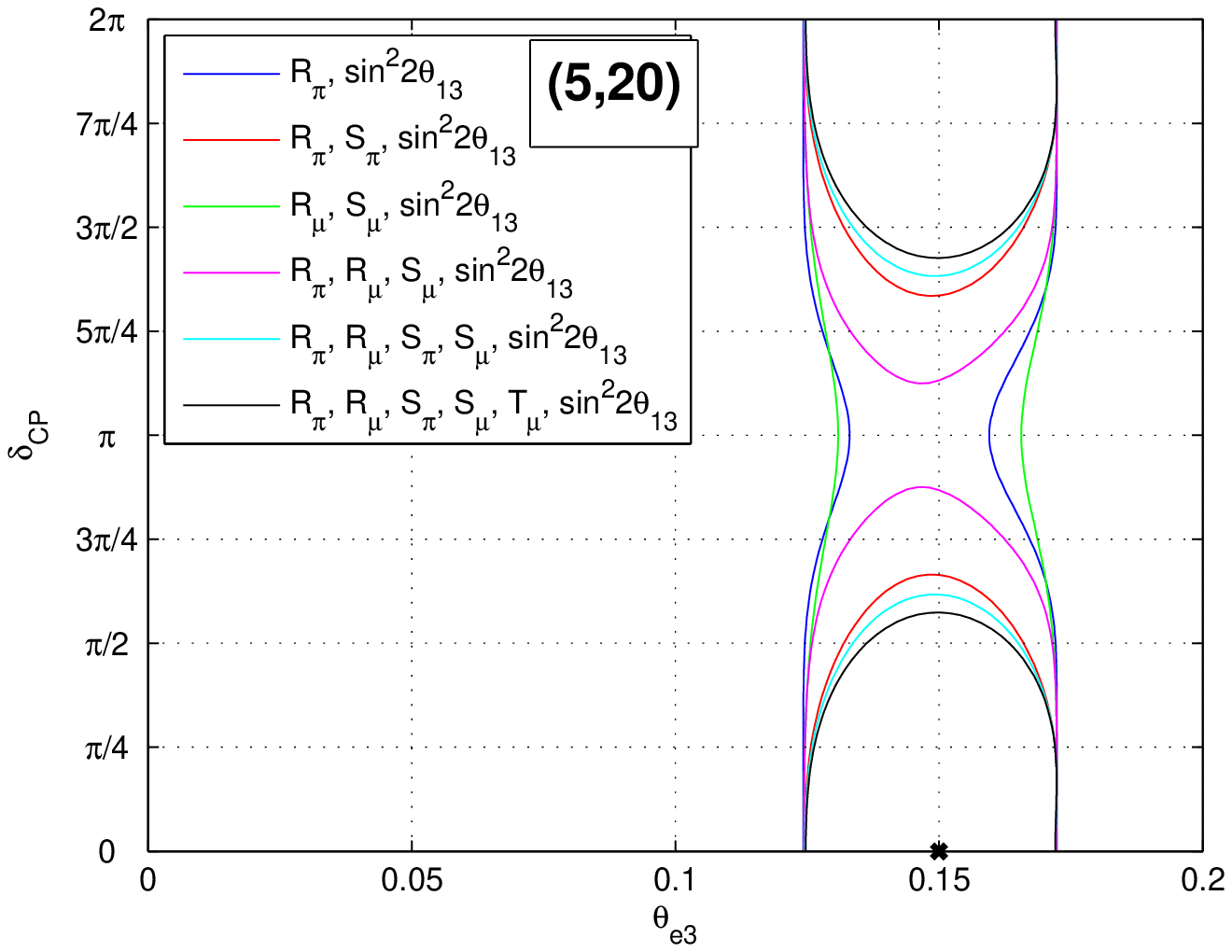} &
\includegraphics[width=80mm,height=60mm]{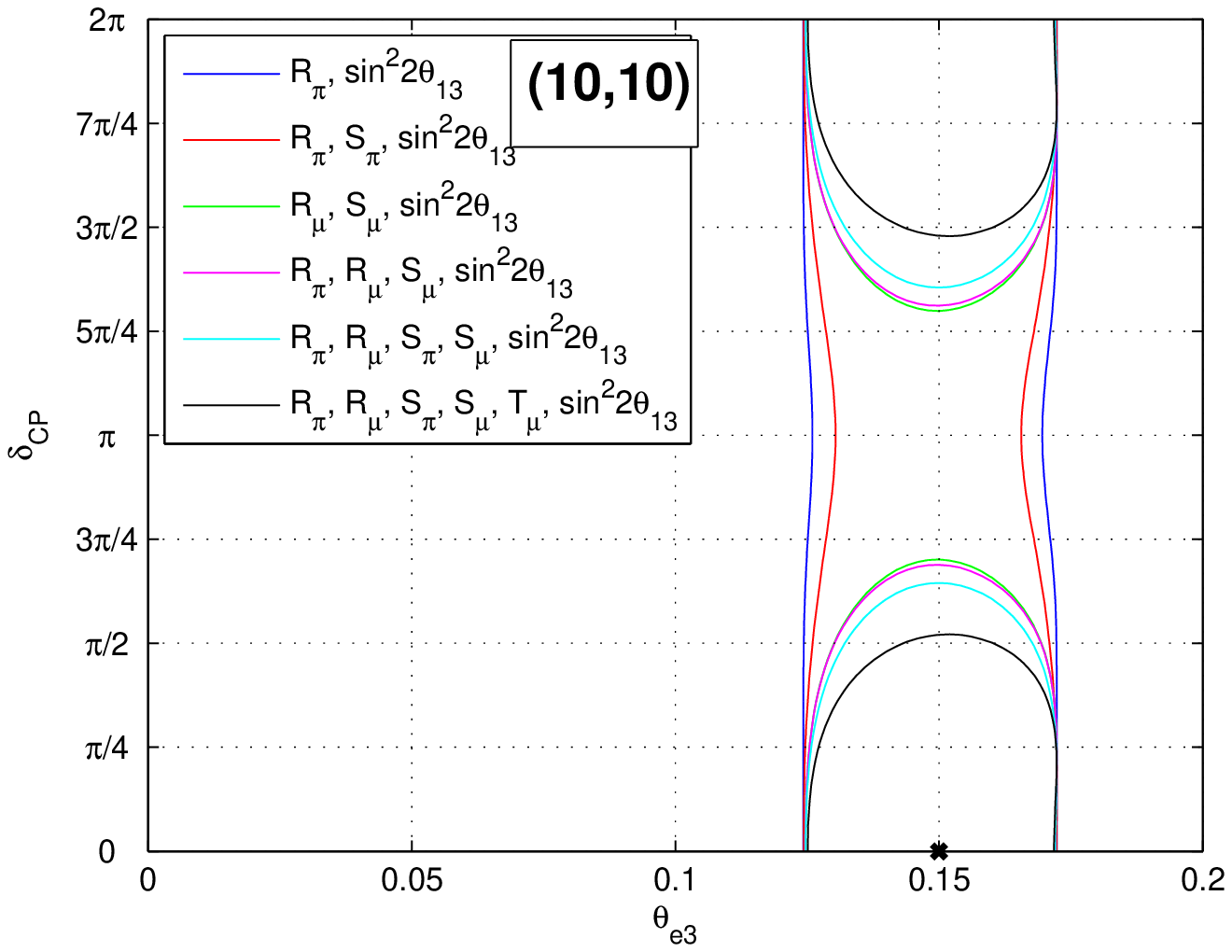} \\
\includegraphics[width=80mm,height=60mm]{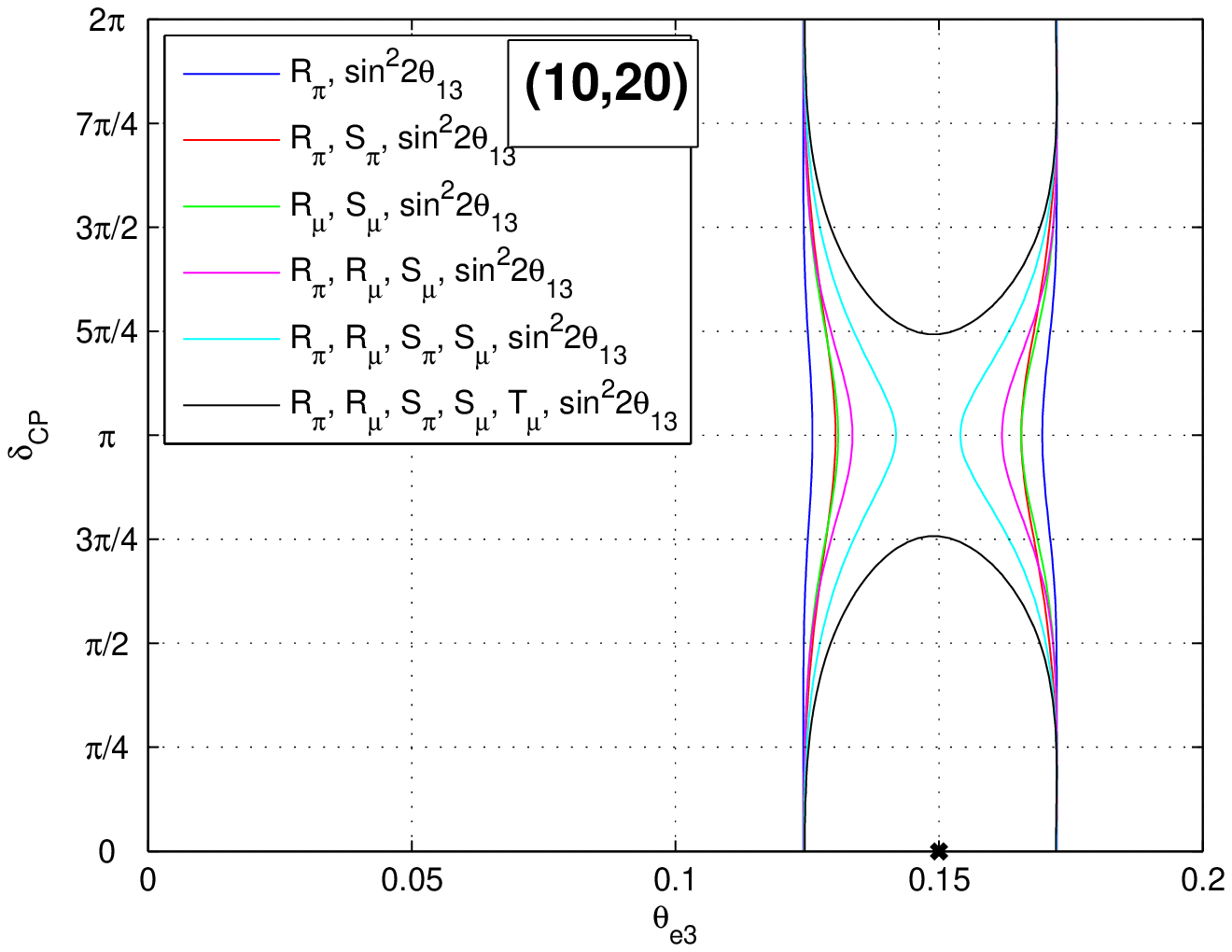} &
\includegraphics[width=80mm,height=60mm]{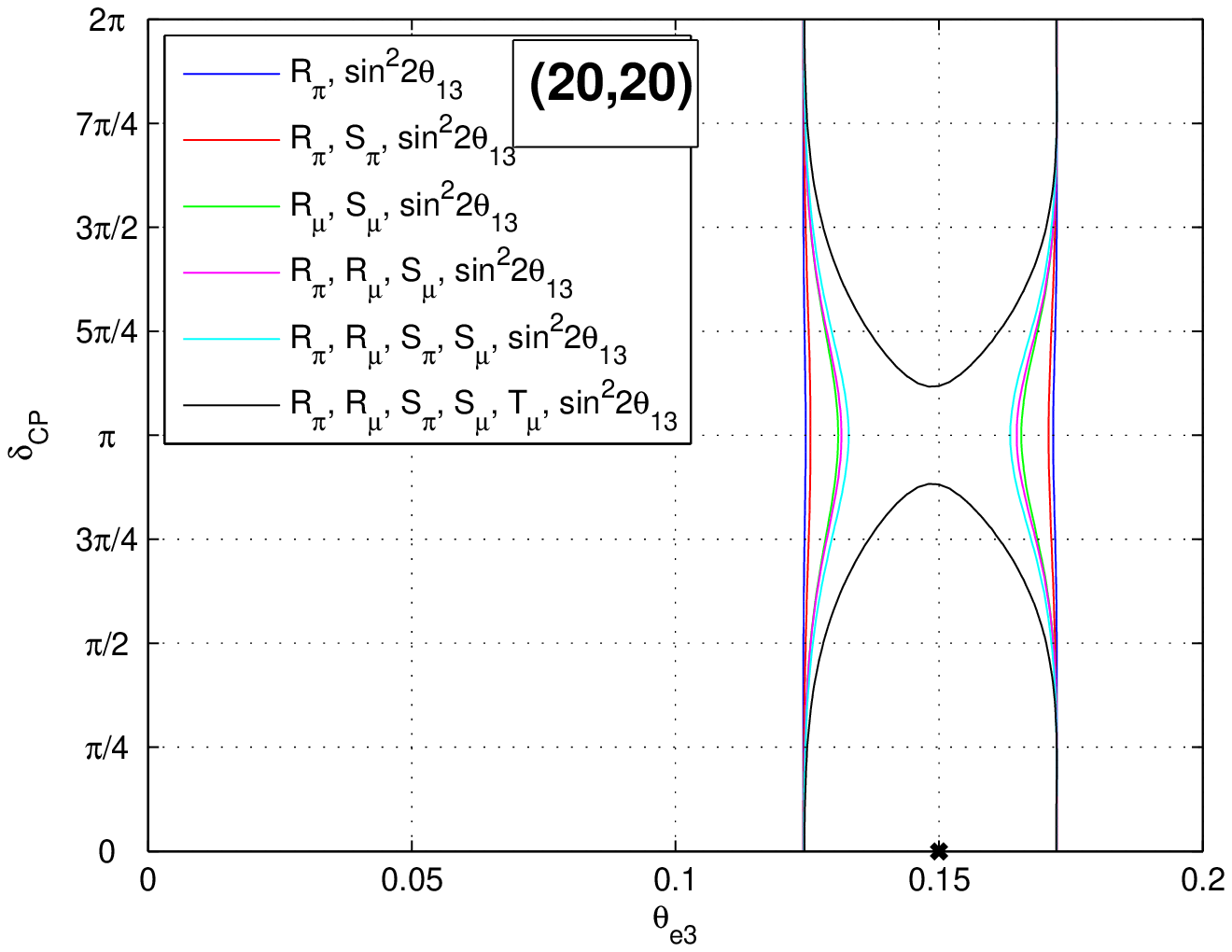}
\end{array}$
\caption{90\%CL (2 d.o.f.) allowed regions for true $\delta_{CP}=0$.}
\label{A1_C1dCP0}
\end{figure}

\begin{figure}[hbp]
\hspace{-2cm} $\begin{array}{lcr}
\includegraphics[width=80mm,height=60mm]{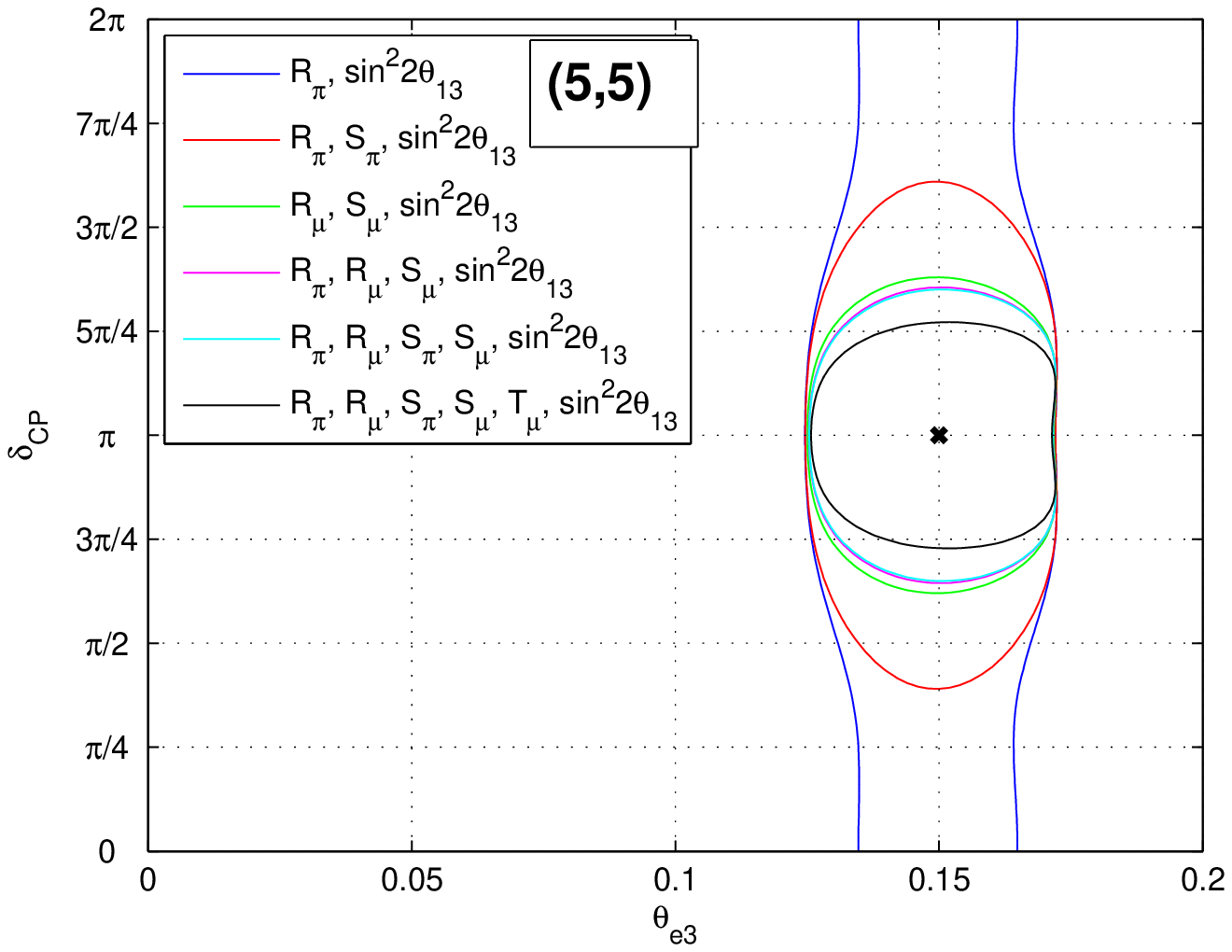}  &
\includegraphics[width=80mm,height=60mm]{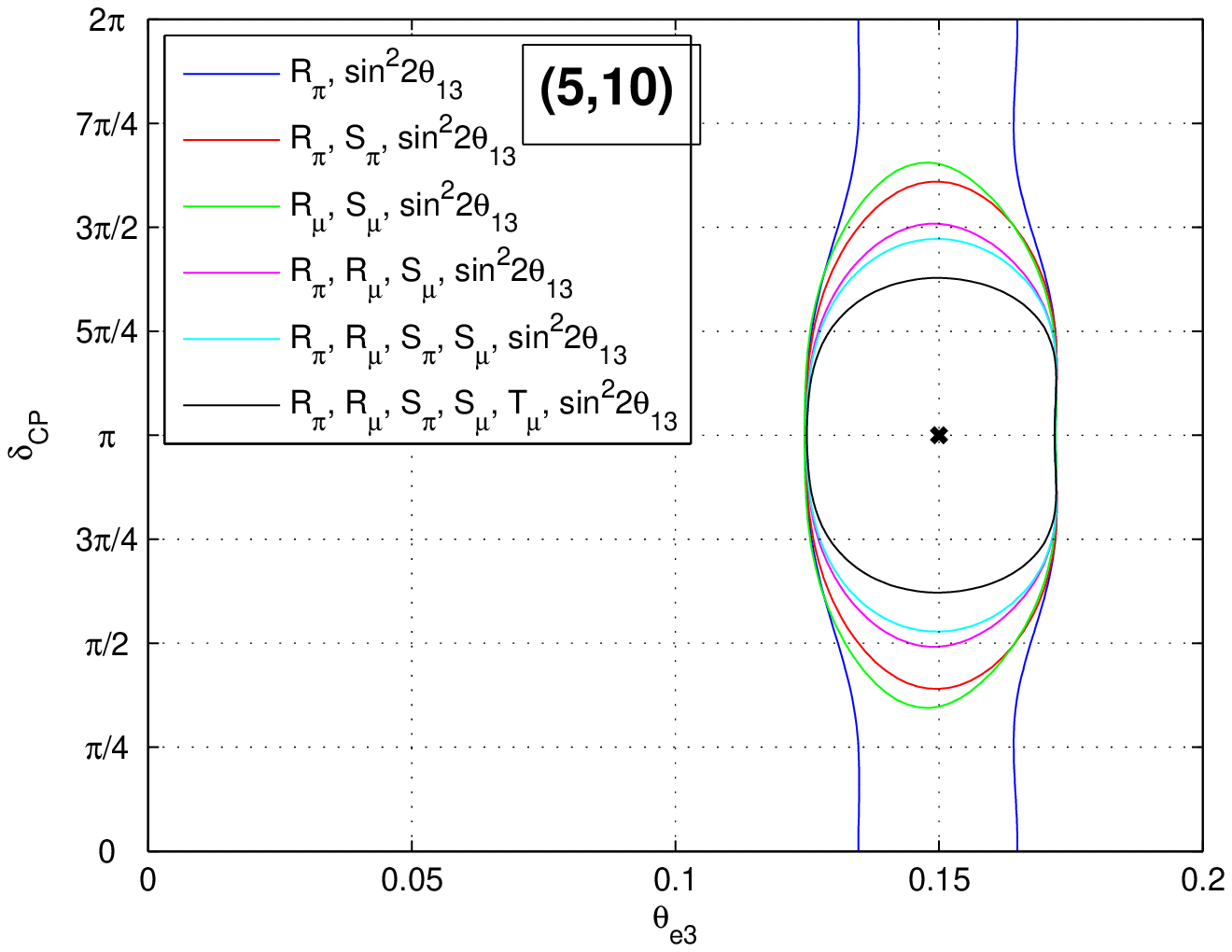}  \\
\includegraphics[width=80mm,height=60mm]{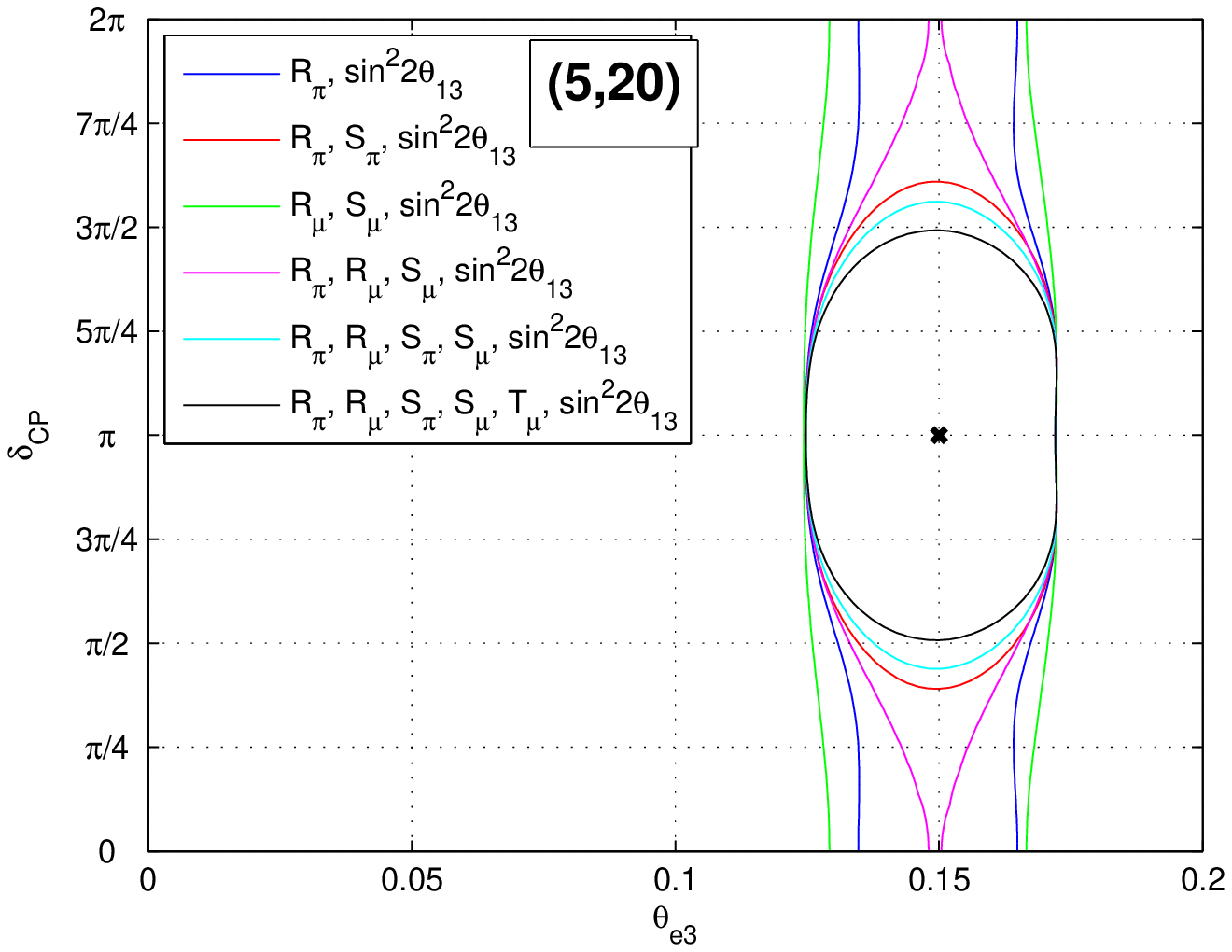} &
\includegraphics[width=80mm,height=60mm]{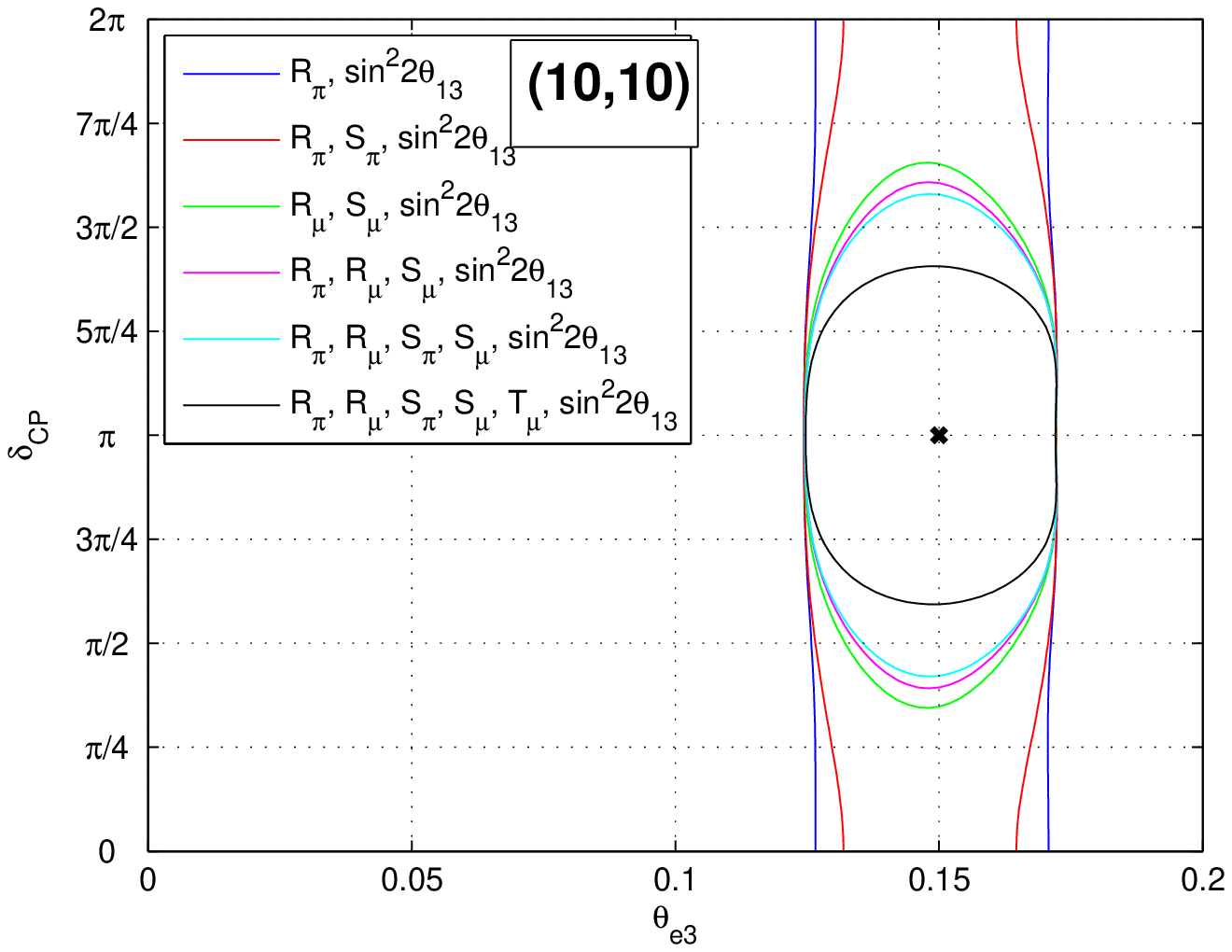} \\
\includegraphics[width=80mm,height=60mm]{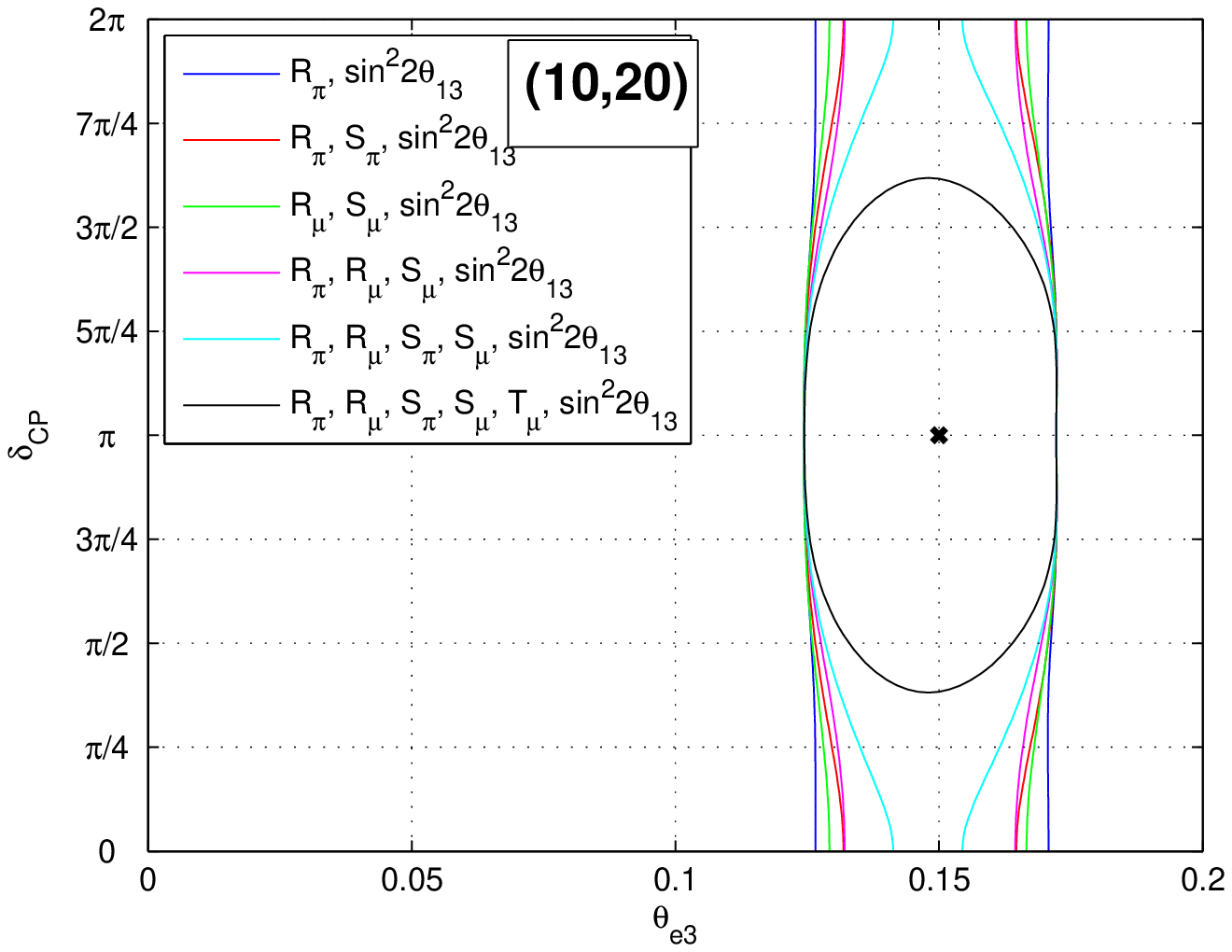} &
\includegraphics[width=80mm,height=60mm]{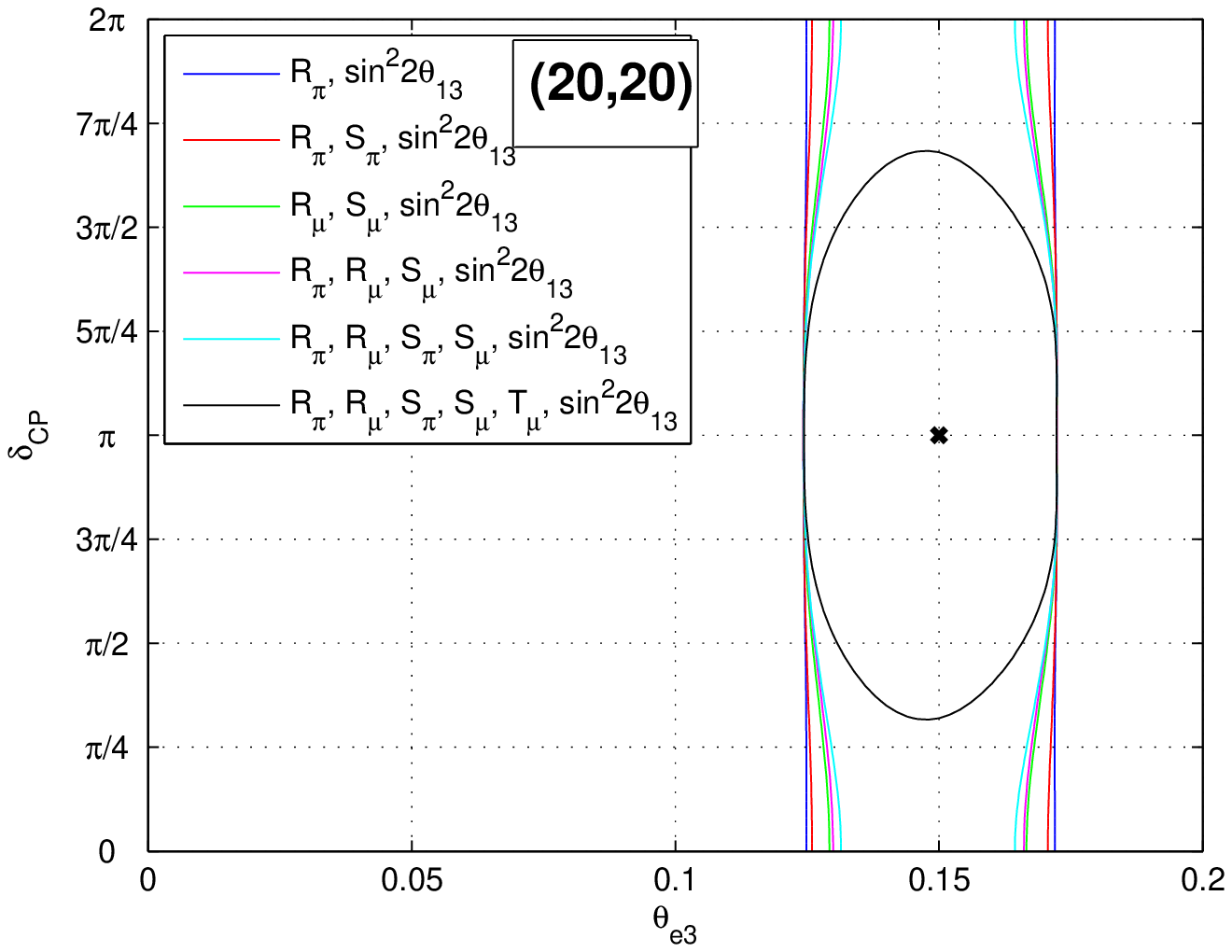}
\end{array}$
\caption{90\%CL (2 d.o.f.) allowed regions for true
$\delta_{CP}=\pi$.} \label{A1_C1dCPpi}
\end{figure}

\begin{figure}[hbp]
\hspace{-2cm} $\begin{array}{lcr}
\includegraphics[width=80mm,height=60mm]{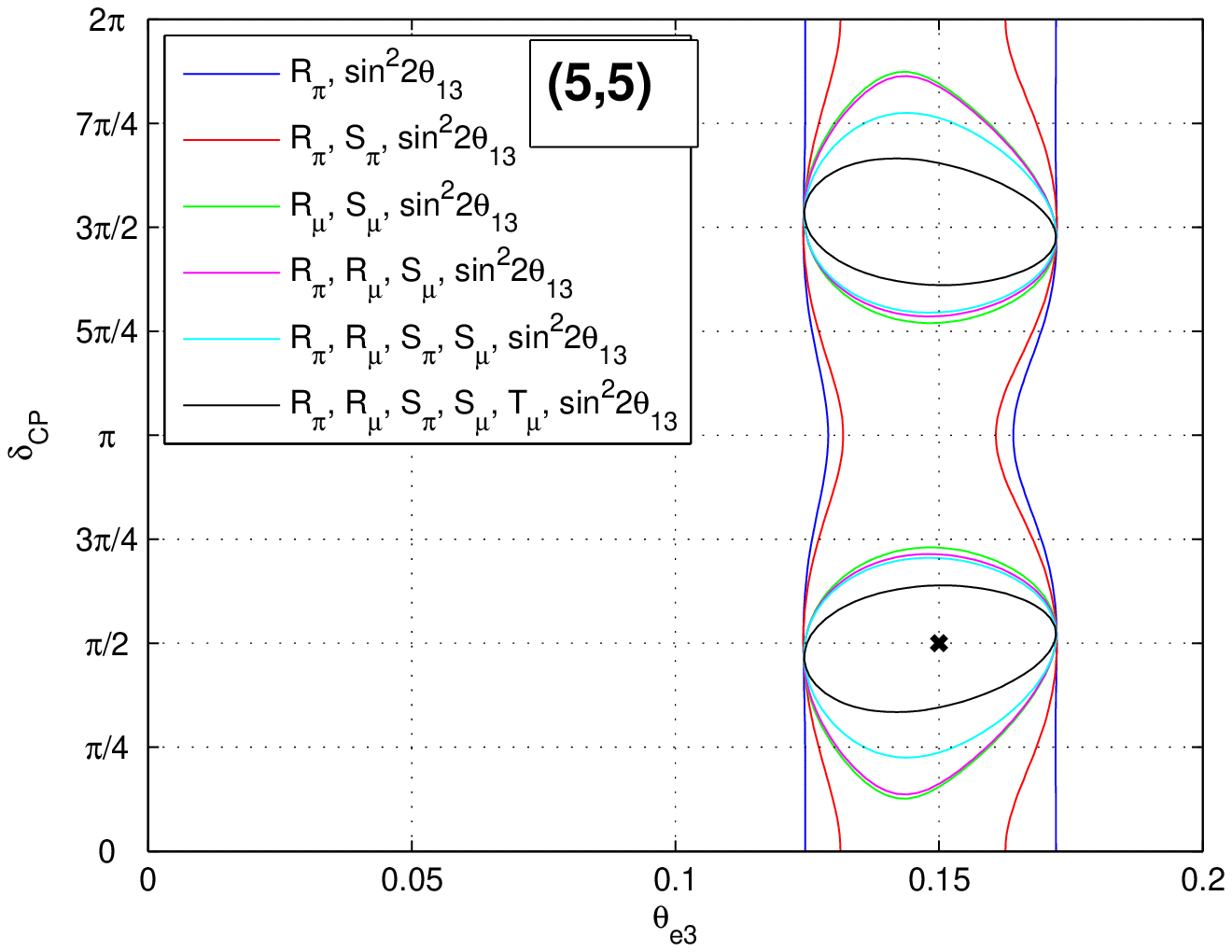}  &
\includegraphics[width=80mm,height=60mm]{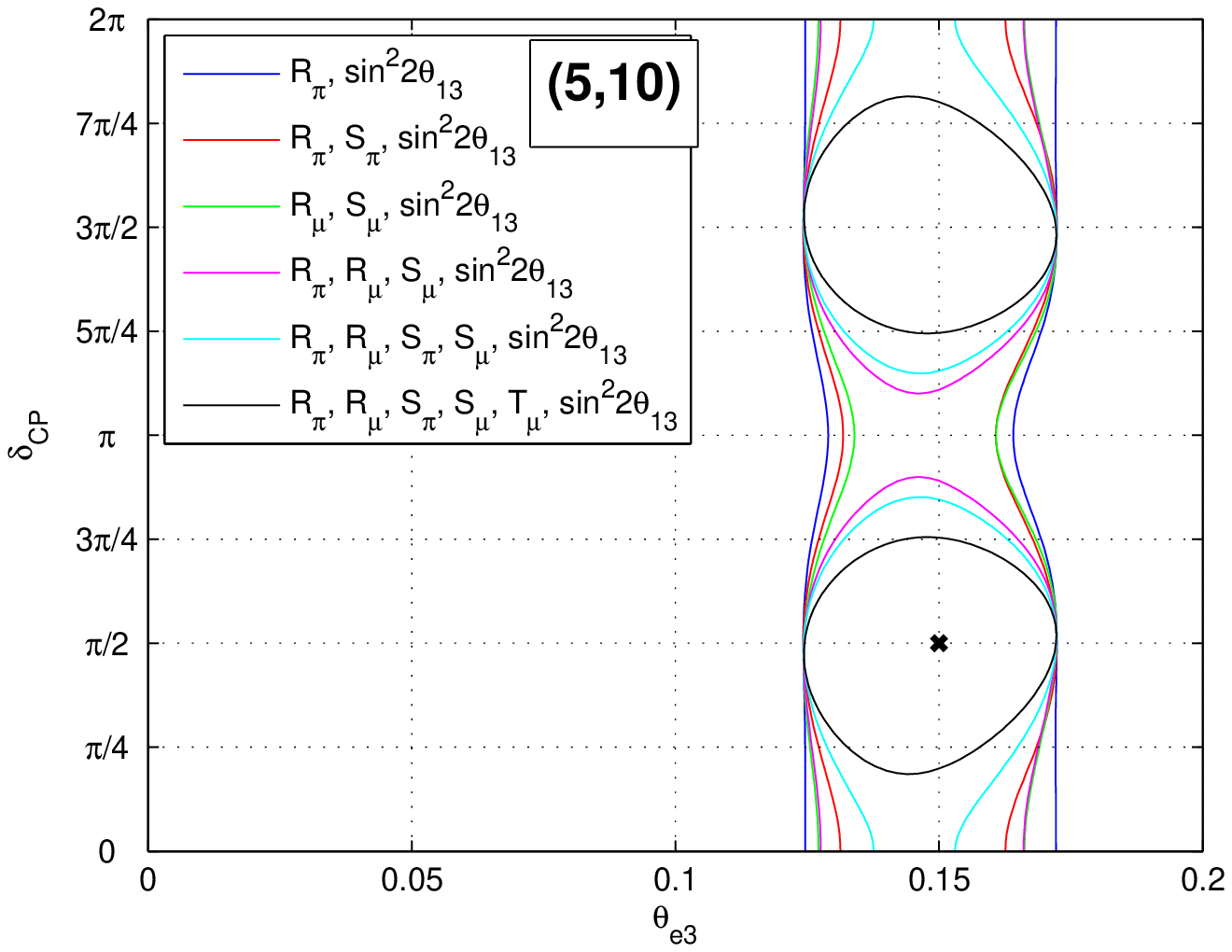}  \\
\includegraphics[width=80mm,height=60mm]{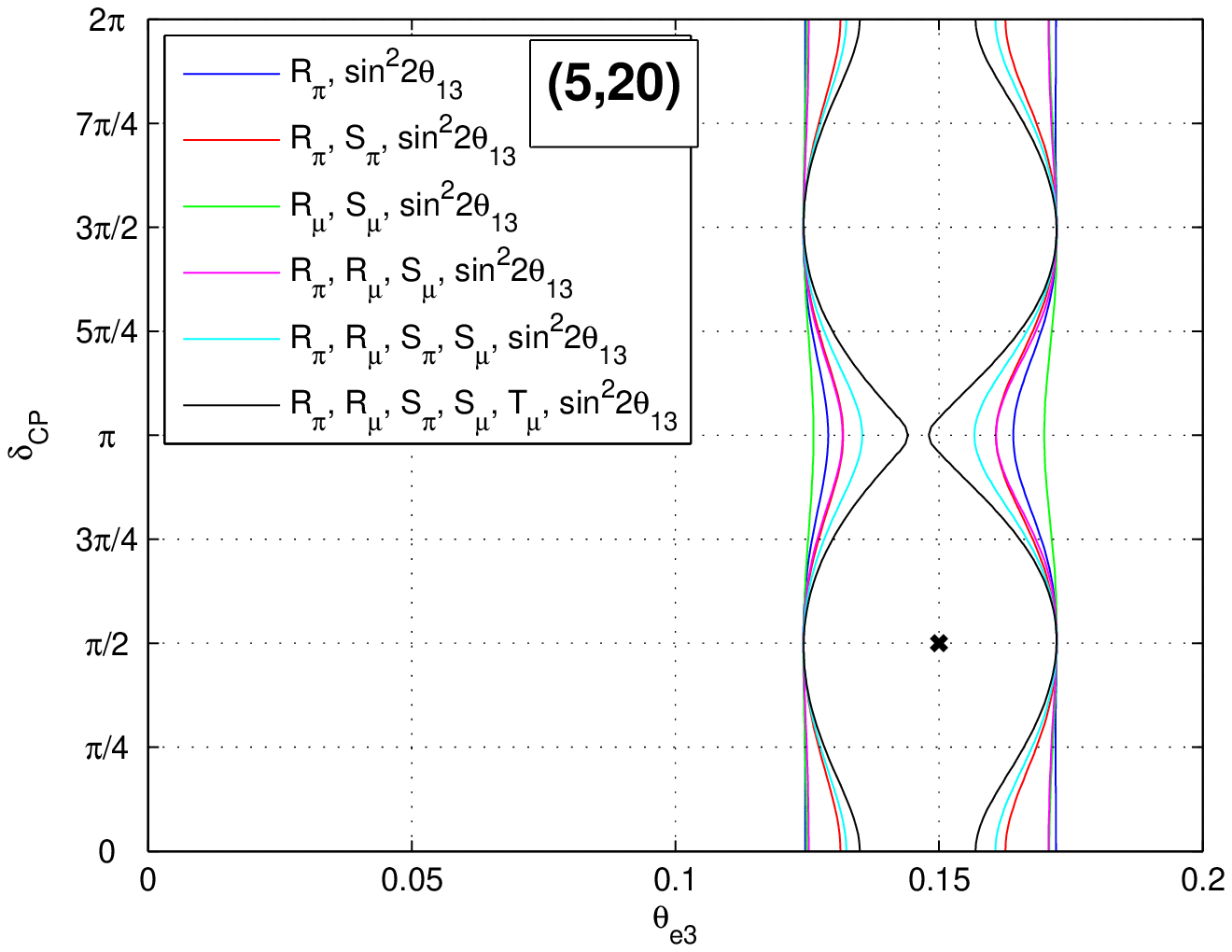} &
\includegraphics[width=80mm,height=60mm]{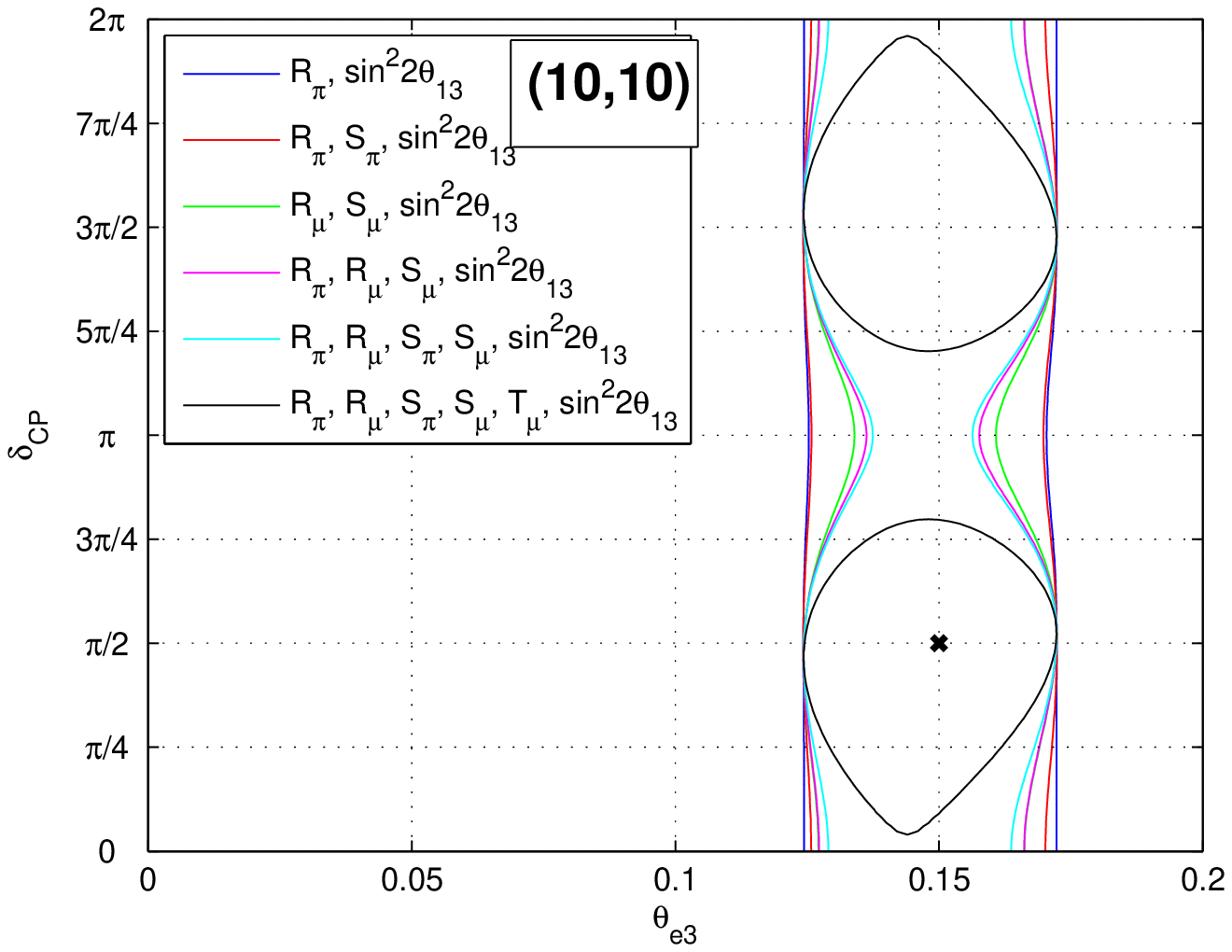} \\
\includegraphics[width=80mm,height=60mm]{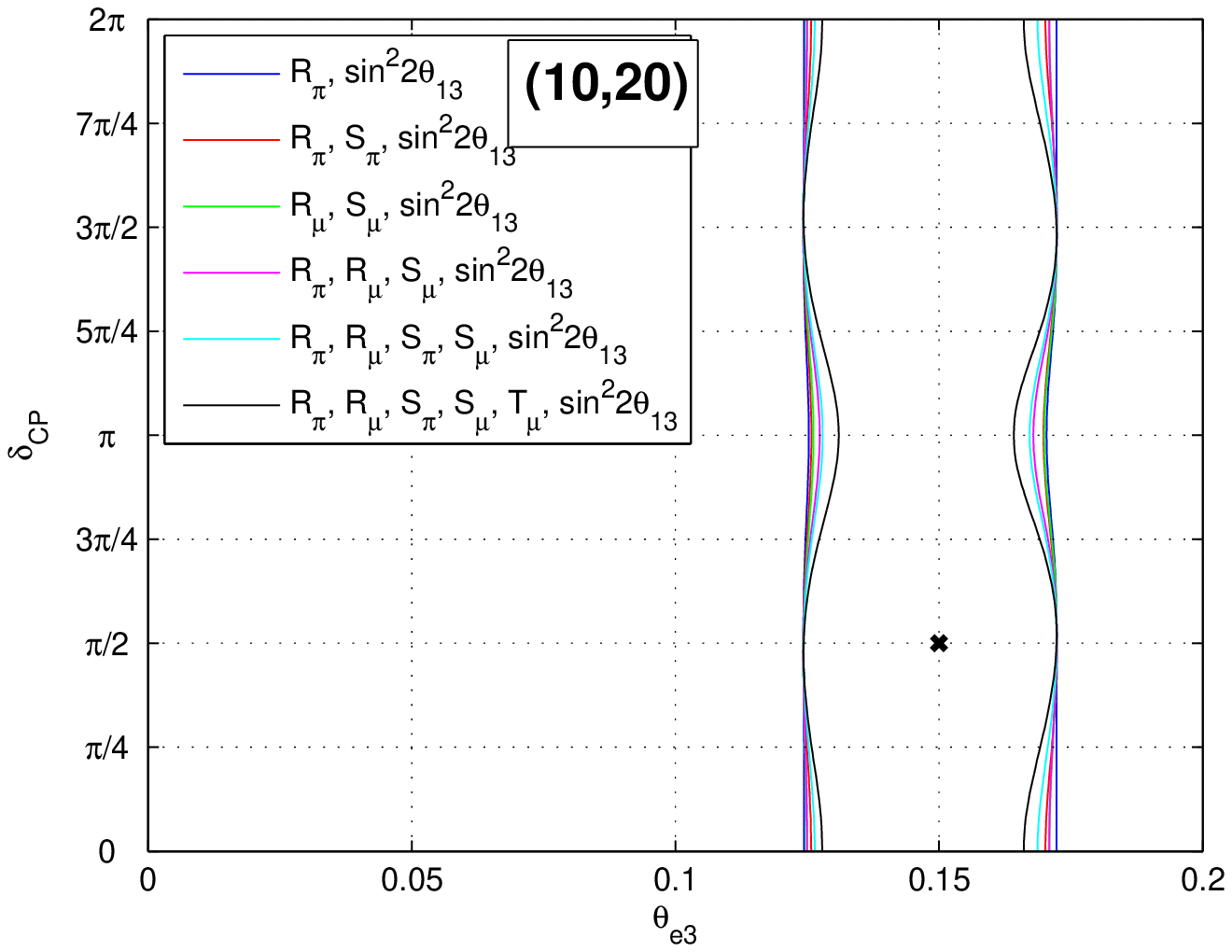} &
\includegraphics[width=80mm,height=60mm]{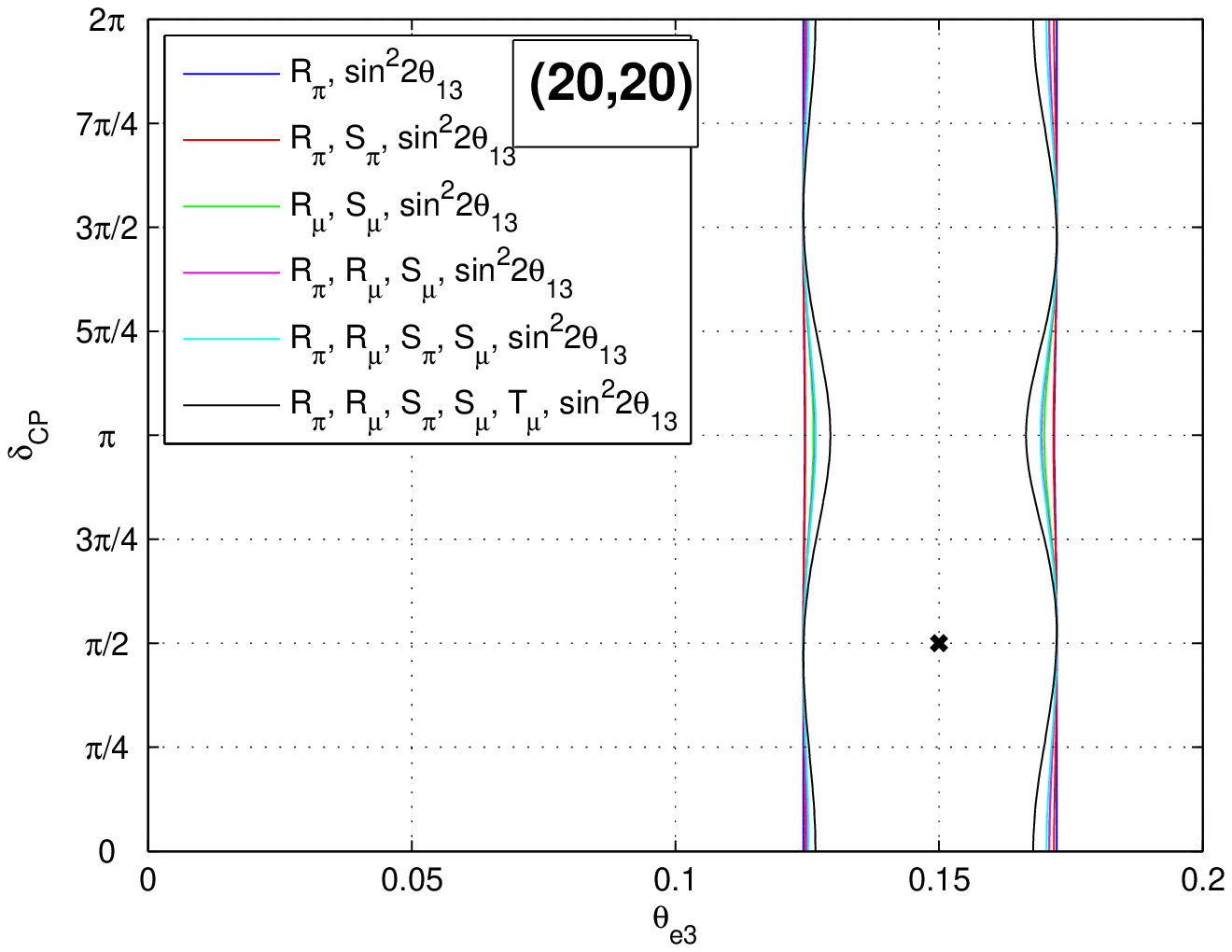}
\end{array}$
\caption{90\%CL (2 d.o.f.) allowed regions for true
$\delta_{CP}=\pi/2$.} \label{A1_C1dCP05pi}
\end{figure}

The resulting 90\% CL acceptance regions in the $\theta_{13}-\delta$
plane are presented, for the six scenarios, in Figs.
\ref{A1_C1dCP0}, \ref{A1_C1dCPpi} and \ref{A1_C1dCP05pi}. As can be
seen in the figures, for some cases, the neutrino telescope
measurements can mildly improve our knowledge of $\theta_{13}$
compared to the reactor constraint.

As concerns $\delta$, the 90\% CL fraction of coverage in case that
{\it all} the relevant observables will be measured is shown in Fig.
\ref{fig:CPfrac}, for true $\theta_{13}=0.15$ and scanning values of
true $\delta$ between $0$ and $\pi$. Since only CP-even quantities
are considered, the results for $\delta=\pi+\theta$ are equal to
those for $\delta=\pi-\theta$. We can make the following statements:
\begin{enumerate}
\item If the neutrino telescope measurements reach the accuracy
  assumed in this work, they are likely to exclude a certain range of
  $\delta$.
\item If the Dirac phase is small (that is close to $0$ or $\pi$), the
  excluded range will be quite significant.
\item The combination of all available observables is usually
  significantly more efficient than partial combinations.
\item The power of combining measurements is particularly significant
  as resolutions get worse and in the large phase ($\delta\sim\pi/2$)
  case.
\item If only $R_\pi$ is measured, no range of $\delta$ will be
  excluded.
\end{enumerate}

The main question that we are asking is the following: Given a
hypothetical situation where $\delta\sim\pi/2$, will IceCube be able
to establish CP violation, that is, exclude 0 and $\pi$ from the
acceptance interval in $\delta$? The answer depends of course on
which of the various scenarios described in Table \ref{tab:models},
if any, will indeed be achieved in the experiment. The main lessons
that we draw from our calculations are the following:

{\bf (5,5):} Measuring $R_\mu$ and $S_\mu$ with an accuracy that is
significantly better than 10 percent will enable a discovery of CP
violation in neutrino oscillations.

{\bf (5,10):} With this scenario, the sensitivity to CP violation is
only marginally affected if either $T_\mu$ or $S_\pi$ are removed
from the analysis. Studying the acceptance interval for $\delta$,
one finds that CP violation may be established even without either
$T_\mu$ or $S_\pi$. This result will be further qualified when we
elaborate on the scenario, below.

{\bf (10,10):} If both $T_\mu$ and $S_\pi$ are measured, with an
accuracy $\sim10\%$, than the required accuracy on $R_\pi$ can be
somewhat relaxed.

{\bf ((5,10,20),20):} If the flavor ratios from muon-damped sources
cannot be measured with an accuracy significantly better than 20\%,
then even an excellent measurement of flavor ratios from pion
sources will not exclude CP conservation.

\begin{figure}[hbp]
\hspace{-1cm} $\begin{array}{lcr}
\includegraphics[width=70mm]{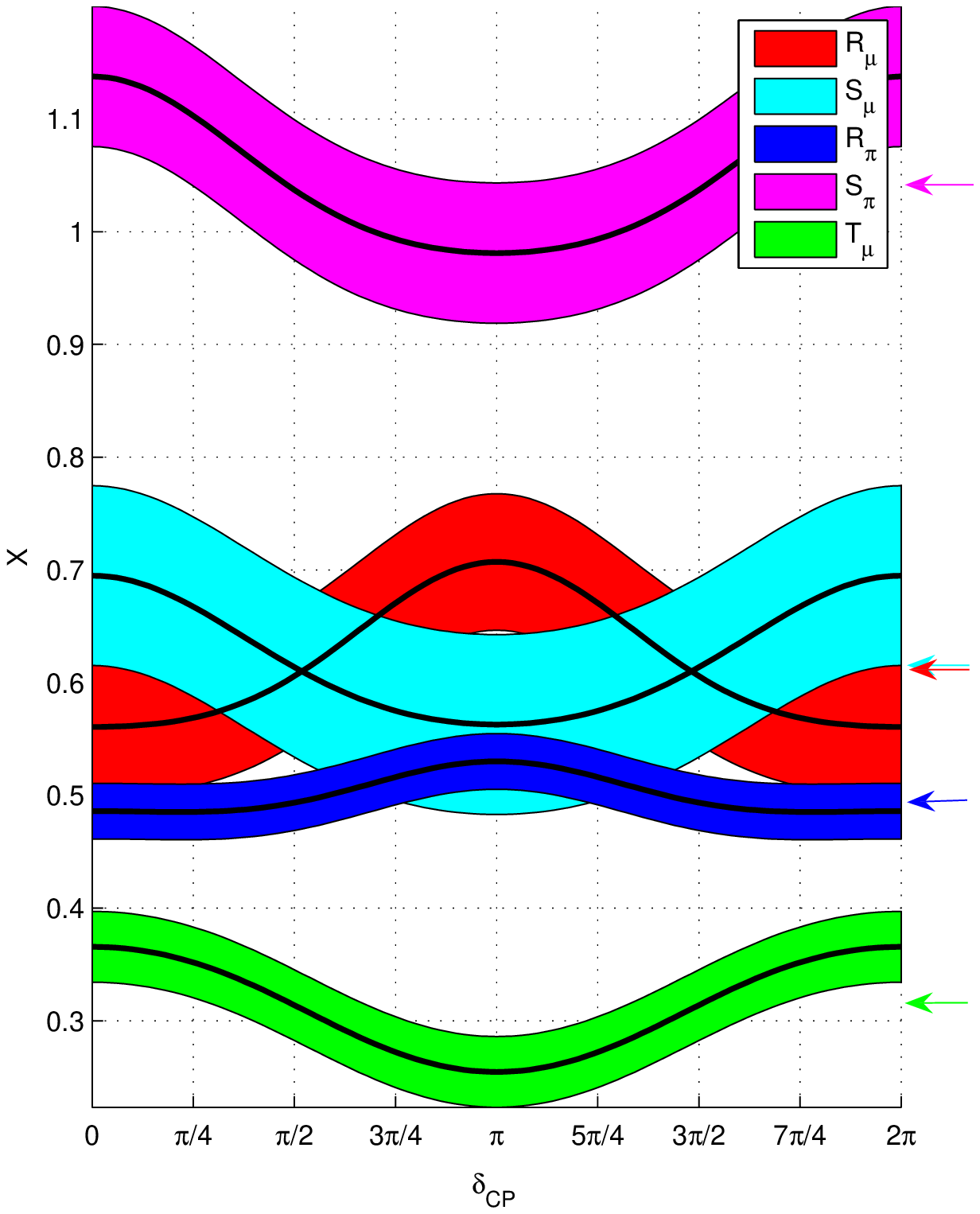}  &
\includegraphics[width=70mm]{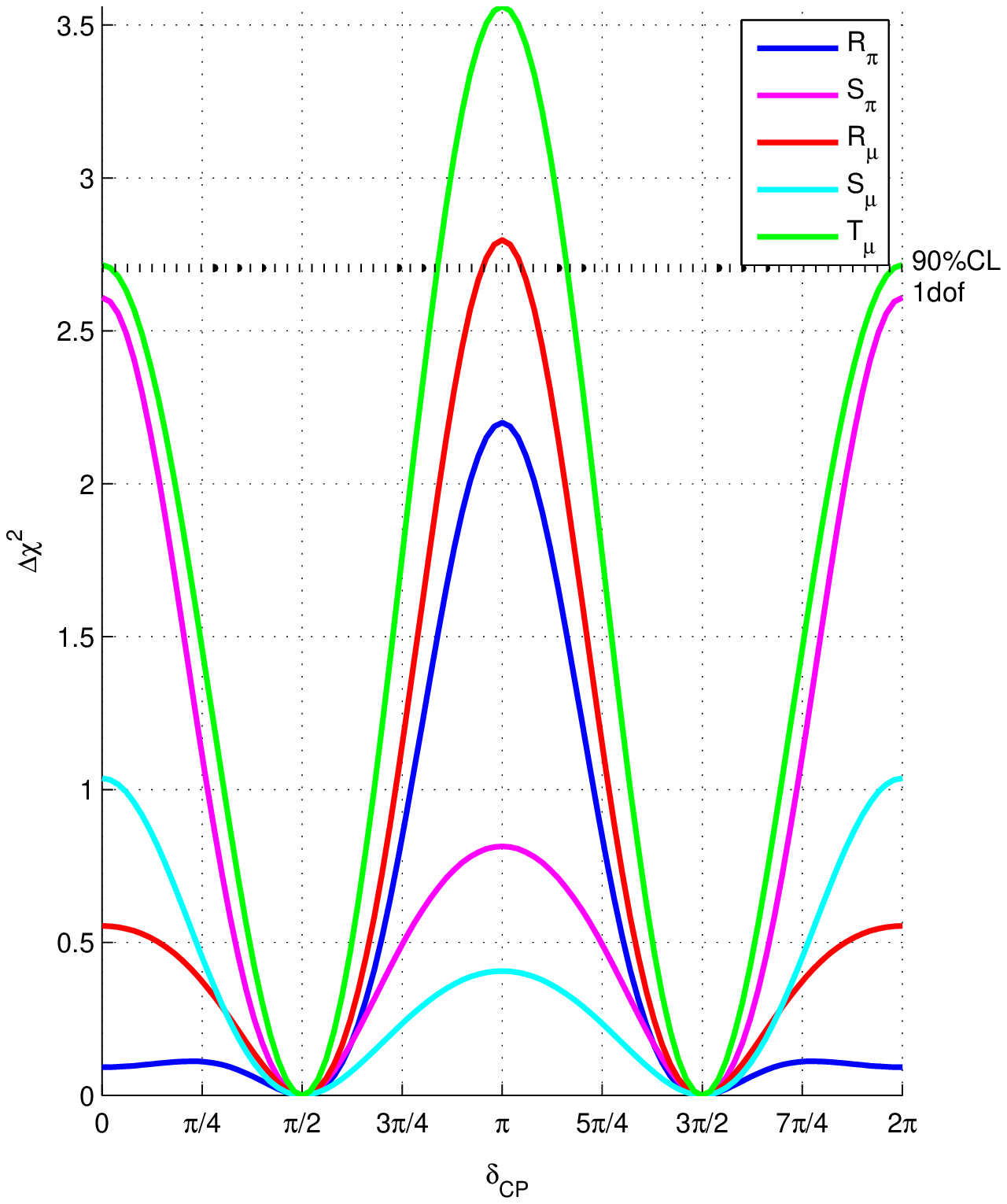}
\end{array}$
\caption{The (5,10) scenario: (left) The flavor ratios as a function
  of $\delta$ and their one-sigma range (arrows mark the central
  values corresponding to $\delta=\pi/2$); (right) The
$\chi^2$ composition for true $\delta=\pi/2$. Both panels correspond
to $\theta_{12}$ and $\theta_{23}$ fixed at their best-fit values,
$\theta_{13}=0.15$.} \label{fig:rstdelta}
\end{figure}

We learn that the $(5,10)$ scenario gives a reasonable sense of the
minimal required set of measurements and accuracies in order that a
discovery that CP is violated in neutrino oscillations will become
possible. Further insight into the role of each of the five
observables in achieving this goal is given in Fig.
\ref{fig:rstdelta}, depicting the flavor ratios as a function of
$\delta$ and the $\chi^2$ composition for true $\delta=\pi/2$. While
measurements of $R_\pi$ and $R_\mu$ at the assumed accuracies
suffice to exclude $\delta=\pi$, at least one of $S_\pi$ or $T_\mu$
needs to be added in order to exclude $\delta=0$.

\begin{figure}[hbp]
$\begin{array}{ccc}
\includegraphics[width=80mm,height=60mm]{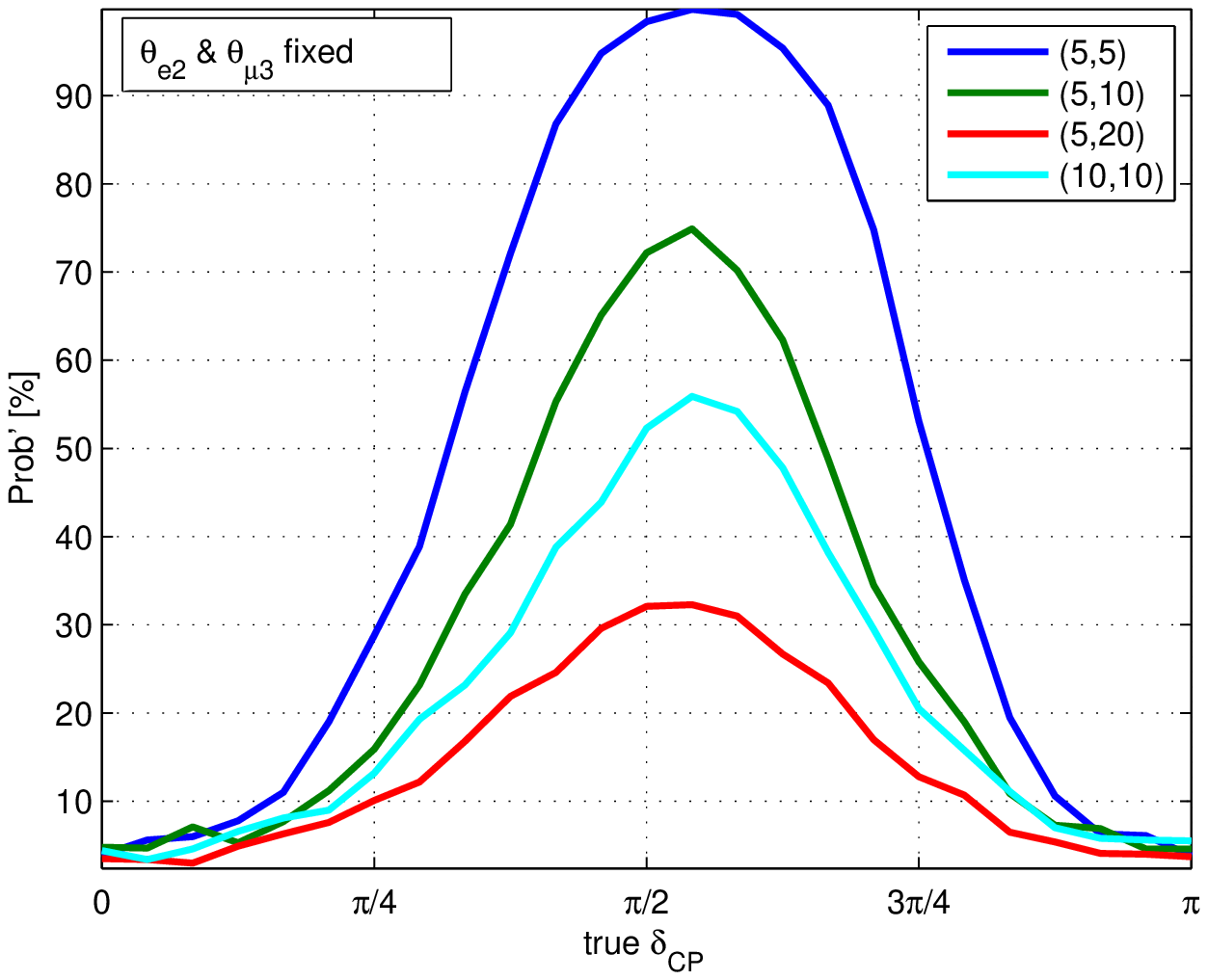}
\includegraphics[width=80mm,height=60mm]{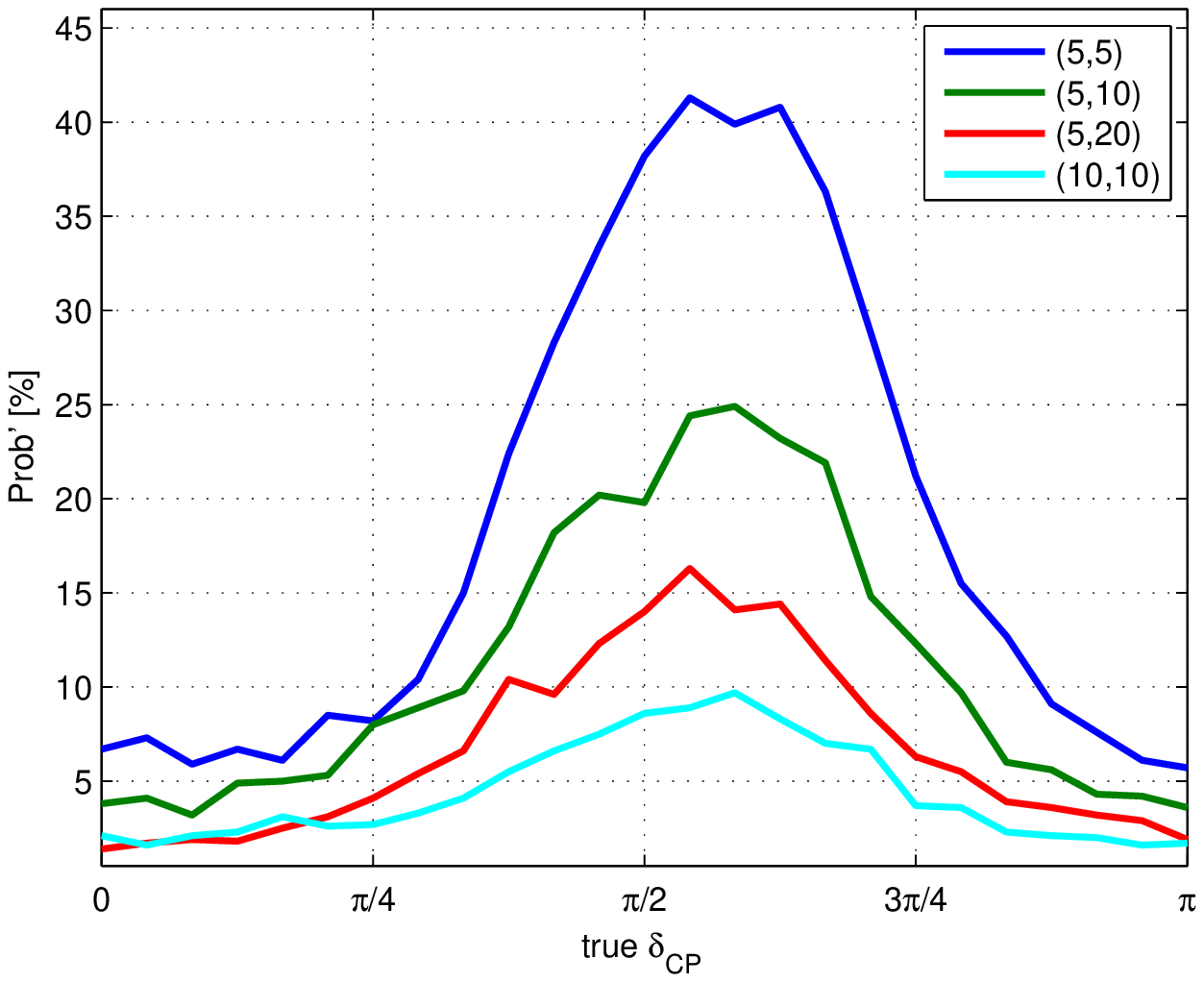}  \\
\includegraphics[width=80mm,height=60mm]{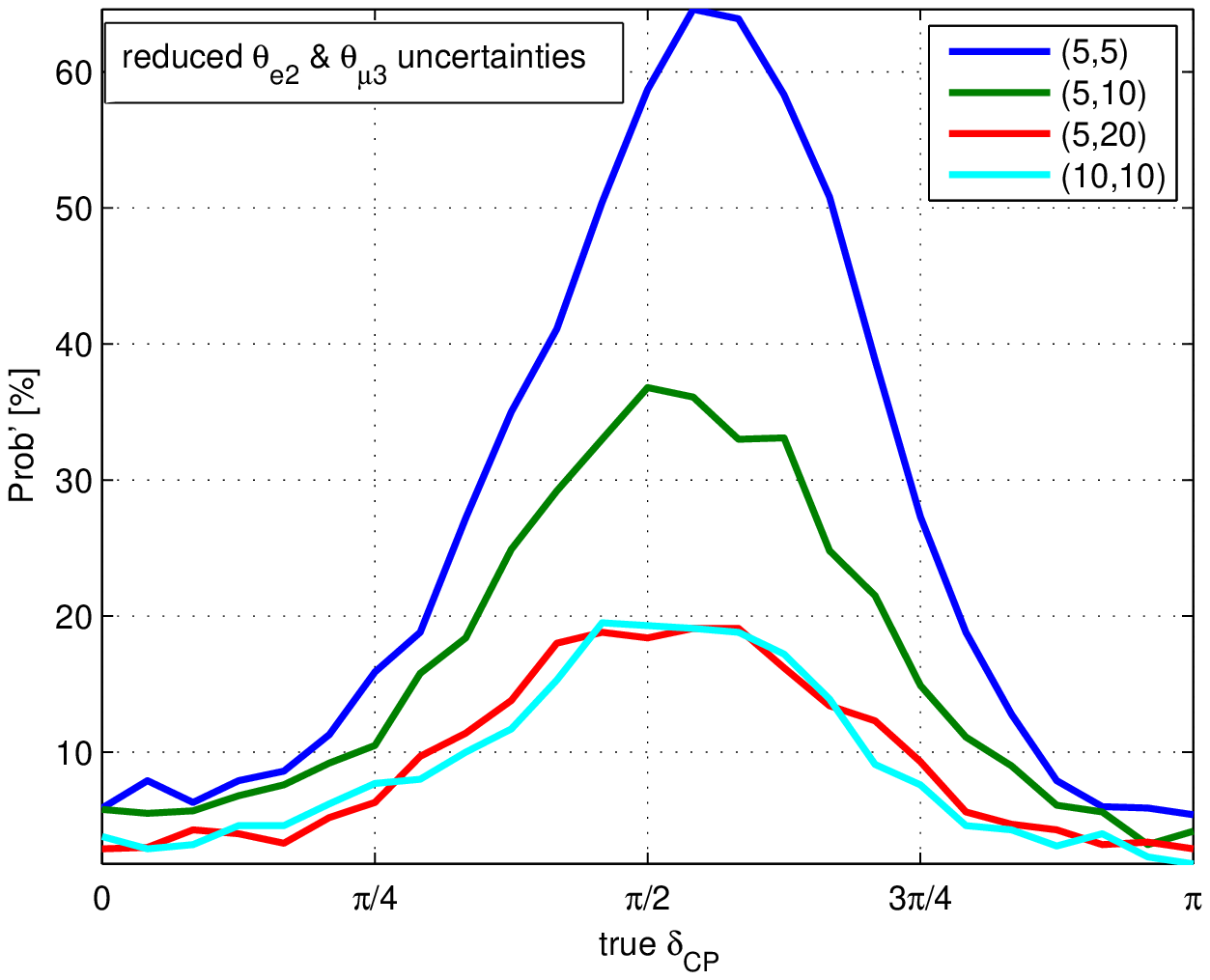}
\end{array}$
\caption{Probability to exclude CP conservation with 90\%CL. The three
  panels differ in the uncertainties attributed to $\theta_{12}$ and
  $\theta_{23}$ (see Table \ref{tab:angles}): (upper left) Zero
  uncertainties; (upper right) Present uncertainties; (bottom)
  `Future' uncertainties.}
\label{fig:cpcexc}
\end{figure}

The probability that CP conserving values of $\delta$ will be
excluded as a function of the true $\delta$, within the four
scenarios (5,5), (5,10), (5,20) and (10,10), is shown in Fig.
\ref{fig:cpcexc}. To produce this plot, we generated a large sample
($1000$) of random sets of observables with the prescribed
statistics, then checked for each realization whether $\delta=0$ or
$\pi$ is contained in the resulting acceptance interval. For
example, with zero uncertainties in $\theta_{12}$ and $\theta_{23}$,
the conditional probability to exclude CP conservation in the
(10,10) scenario given maximal phase is about 50\%. Note that
statistical fluctuations may lead to erroneous exclusion of CP
conservation even with $\sin\delta=0$. The fact that the (10,10)
scenario is more likely than (5,20) to establish CP violation is
suggestive for future detector optimizations: If the errors on
$\theta_{12}$ and $\theta_{23}$ at the time of analysis are
significantly reduced, then it may be preferable to improve the
detection efficiency at the higher range of the spectrum, $E>100\
TeV$, even at the cost of somewhat weaker efficiency at lower
energies.

%%%%%%%%%%%
\subsection{Taking into account uncertainties in $\theta_{12}$ and
  $\theta_{23}$}
\begin{figure}[hbp]
$\begin{array}{ccc}
\includegraphics[width=80mm,height=60mm]{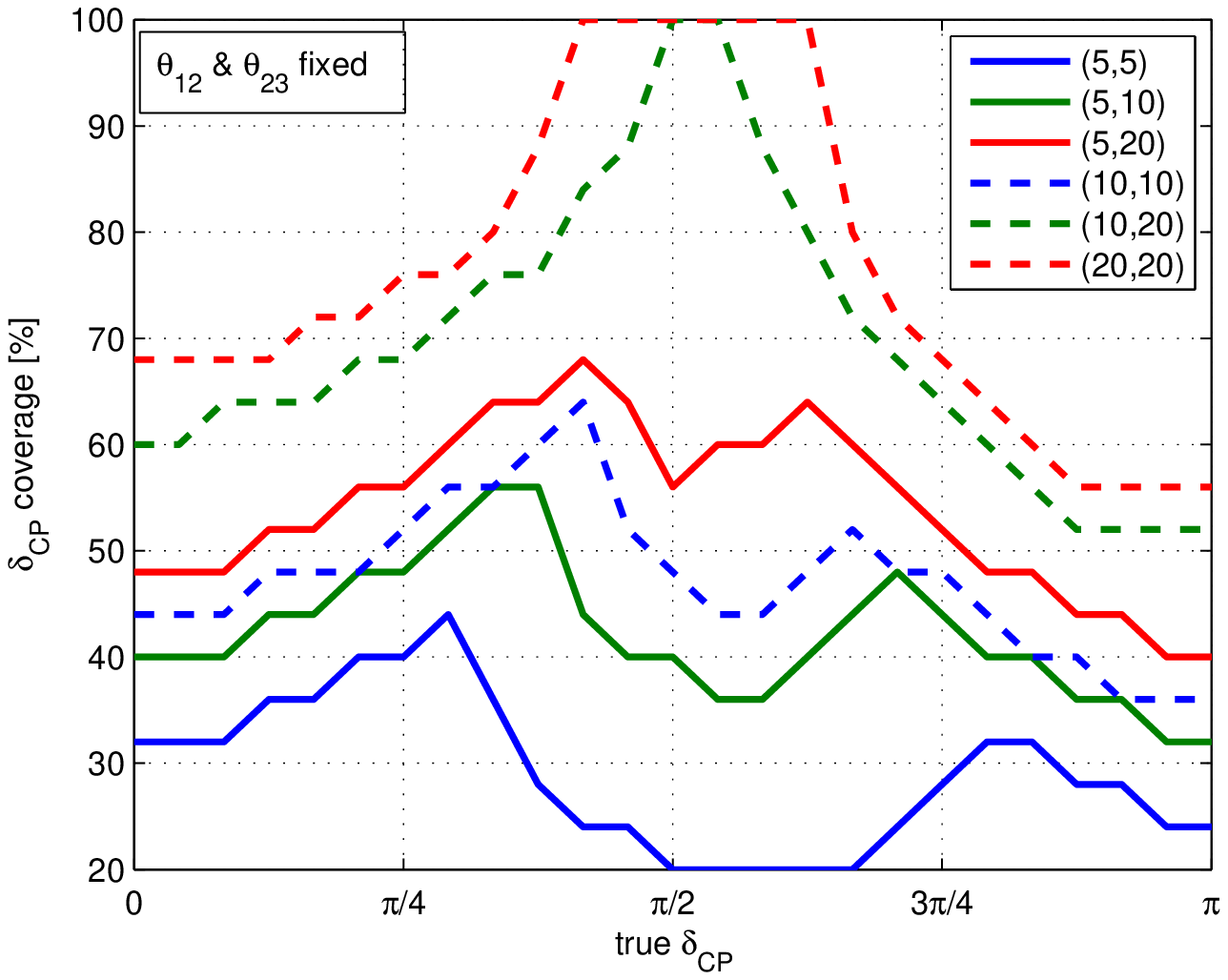}
\includegraphics[width=80mm,height=60mm]{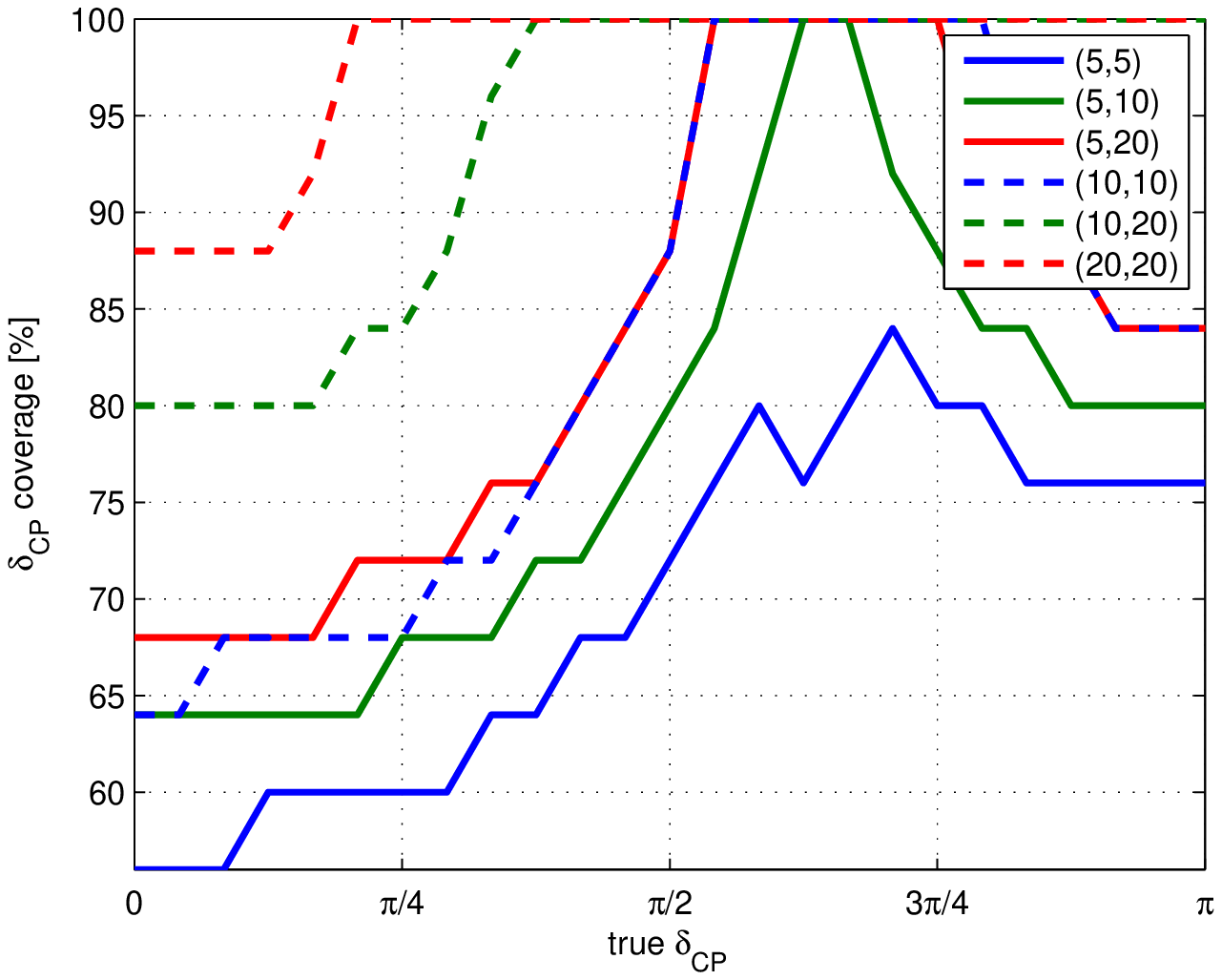}  \\
\includegraphics[width=80mm,height=60mm]{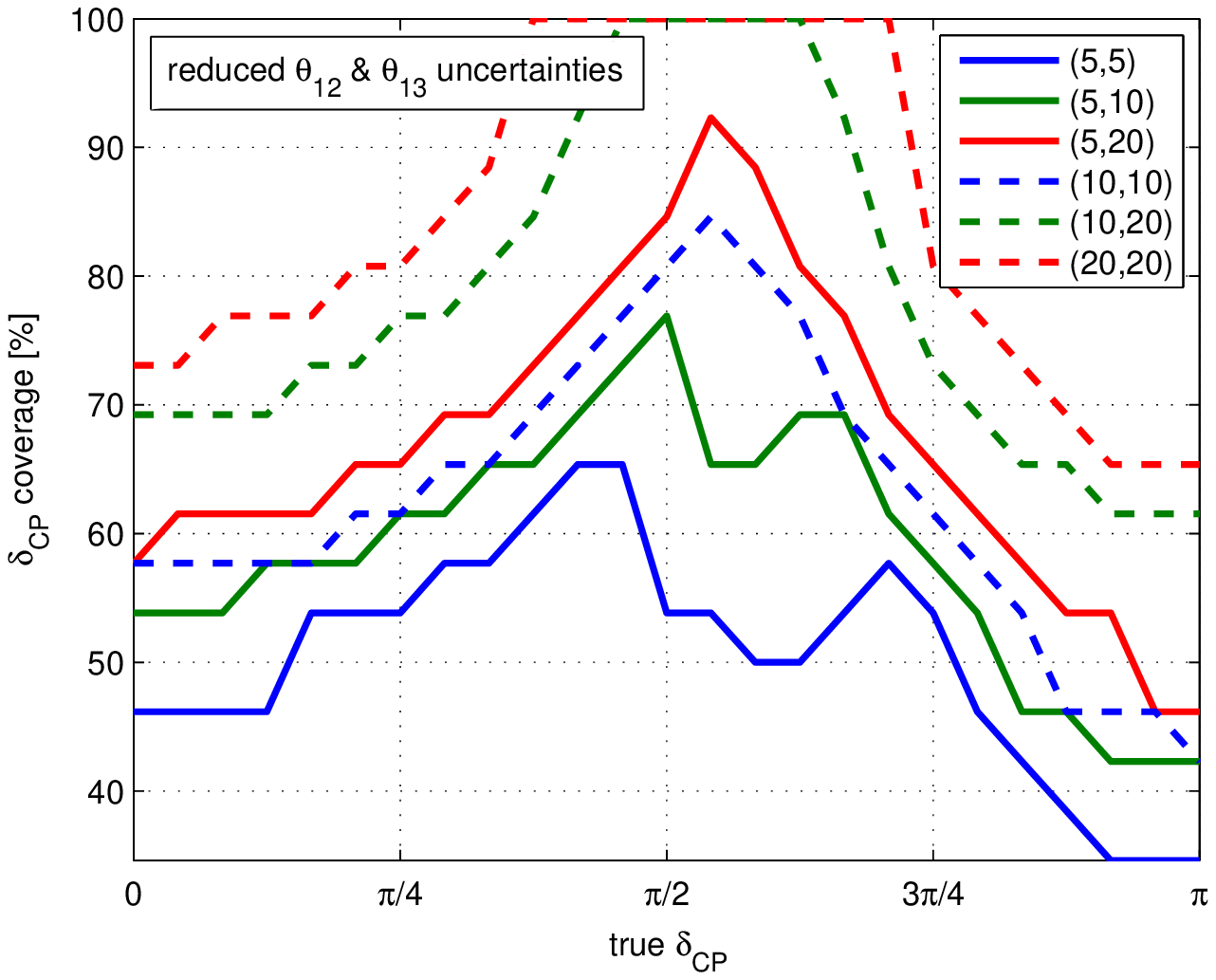}
\end{array}$
\caption{90\% CL fraction of coverage for $\delta$ in the six scenarios defined
in Table \ref{tab:models}. The three panels differ in the
uncertainties attributed to $\theta_{12}$ and $\theta_{23}$ (see
Table \ref{tab:angles}): (upper left) Zero uncertainties; (upper
right) Present uncertainties; (bottom) `Future' uncertainties.} \label{fig:CPfrac}
\end{figure}

As a first step in this analysis, we considered the present ranges
for $\theta_{12}$ and $\theta_{23}$ (see Table \ref{tab:angles}).
The potential of neutrino telescopes to exclude a range of $\delta$
can be seen from Fig. \ref{fig:CPfrac} (upper right panel). The
impact of the uncertainties in $\theta_{12}$ and $\theta_{23}$ can
be understood by comparing it to the upper left panel. We learn
that, with present accuracies, the excluded ranges are weaker by
30-50\% compared to the idealized case of zero uncertainties. (The
importance of this ingredient in the analysis was noted in
\cite{Meloni:2006gv}.)

As a second step, we assumed experimental errors on
$\sin^2\theta_{12}$ and $\sin^2\theta_{23}$ that are reduced by a
factor of two compared to the present (see Table \ref{tab:angles}).
The results are shown in Fig. \ref{fig:CPfrac} (lower panel). By
comparing to the upper right panel, we learn that such an improvement
will entail an exclusion power stronger by about 20\% compared to the
situation that present uncertainties remain.

Concerning the probability that CP violation will be established, we
repeat the analysis with the present and with the assumed future
uncertainties for the four leading scenarios. The results are shown
in Fig. \ref{fig:cpcexc}. Without an improvement in the
determination of $\theta_{12}$ and $\theta_{23}$, only the very
optimistic scenario (5,5) allows a discovery. With the assumed
improvements, the more realistic (5,10) scenario also has over
$30\%$ probability to make such a discovery. The (5,20) and (10,10)
scenarios are not powerful enough to do so.

%%%%%%%%%%%%%%%%%
\subsection{Discussion}
A related analysis has been performed previously in
Refs. \cite{Winter:2006ce,Balaji:2006wi}, which highlighted the synergy
between neutrino telescopes and terrestrial experiments. The
conclusion in Refs. \cite{Winter:2006ce,Balaji:2006wi} regarding the
impact of neutrino telescopes on the issue of CP violation is more
pessimistic than ours. The main difference lies in the fact that
Refs. \cite{Winter:2006ce,Balaji:2006wi} 
consider the information of one type of sources at a time, and indeed
we agree with the pessimistic conclusion in this case. What we show,
however, is that by combining the two types of sources that we
considered, the ability to exclude CP conservation improves
considerably. Actually, if this combination of sources is indeed
available (and the experimental accuracy is similar to or better than
our (10,10) scenario), the exclusion power that neutrino telescopes
have on $\delta$ will be comparable to the proposed superbeams
\cite{Huber:2004ug}. (This situation actually reinforces the point
made in \cite{Winter:2006ce}: since the $\delta$-dependencies of the
IceCube and the superbeam measurements are different, the information
from the two will be complimentary.)

Ref. \cite{Lipari:2007su} points out that variations in the flavor
ratios between sources can reach the ten percent level and
consequently play an important role in the investigation of the
mixing parameters from astrophysical neutrinos. In particular, the
resulting uncertainties may wash-out the effects of the
$\Delta_{13}$ terms, especially in the case of low $\theta_{13}$. We
agree that flavor composition uncertainties at the source would
tighten greatly the requirements on the experimental precision.
There are two reasons, however, why we think that this issue may
have only limited consequences for our purposes. First, by the time
that this analysis can be carried out in IceCube, the theoretical
analysis of neutrino spectra, which is only at its beginning
\cite{Kashti:2005qa,Lipari:2007su}, is likely to improve
considerably. In particular, higher quality electromagnetic data,
from radio to TeV photon energies, will become available. Second,
our study is relevant only for the case of large $\theta_{13}$
where, as we have argued, $10\%$ accuracy might be just enough for
our purposes if a global analysis of flavor-dependent spectrum will
be possible.

The general trends reflected in our results can be simply understood,
based on Eqs. (\ref{piobs}) and (\ref{muobs}). We rewrite them as
follows:
\beqa\label{pimuobs}
R_\pi&=&0.49\left[1-0.05(s_{13}/0.15)\cos\delta\right],\no\\
S_\pi&=&1.04\left[1+0.08(s_{13}/0.15)\cos\delta\right],\no\\
R_\mu&=&0.62\left[1-0.12(s_{13}/0.15)\cos\delta\right],\no\\
S_\mu&=&0.58\left[1+0.11(s_{13}/0.15)\cos\delta\right],\no\\
T_\mu&=&0.30\left[1+0.19(s_{13}/0.15)\cos\delta\right].
\eeqa
We learn the following:
\begin{itemize}
\item The ratios related to muon-damped sources are more sensitive to
  the $\cos\delta$-dependent terms than those related to pion sources;
\item To be sensitive to the $\cos\delta$-dependent terms, the
  accuracy should be of order 10\% or better;
\item The required accuracy scales with $s_{13}$. If, for example,
  $s_{13}\sim0.05$, sensitivity to $\cos\delta$ will be achieved only
  with accuracy better than 5\%, which seems out of reach for IceCube.
\end{itemize}

%%%%%%%%%%%%%%%%%%%%%%%
\section{Conclusions}
\label{sec:con}
We have studied the potential of combining measurements of flavor
ratios in neutrino telescopes with observation of $\theta_{13}\neq0$
by reactor experiments in constraining $\delta$, the CP violating
phase in the lepton mixing matrix. We reached the following
conclusions:
\begin{itemize}
\item Since the neutrino telescopes are sensitive only to the
  combination $\Delta_{13}\equiv\sin\theta_{13}\cos\delta$, they can
  constrain $\delta$ only if $\sin\theta_{13}$ is not too small
  \cite{Winter:2006ce}.
\item Neutrino telescope may exclude at 90\% CL up to 30\% of the
  a-priori allowed range for $\delta$, even with present accuracies in
  $\theta_{12}$ and $\theta_{23}$.
\item Since the $\Delta_{13}$-term is maximized in size for
  $\cos\delta=\pm1$, the exclusion region is largest if CP is nearly
  conserved \cite{Winter:2006ce}.
\item Reduced uncertainties in $\theta_{12}$ and $\theta_{23}$ can
  enlarge the excluded region to about 50\% of the a-priori allowed
  range, and give sensitivity even for $\cos\delta\sim0$.
\item Measuring flavor ratios of fluxes from muon-damped sources will
  further strengthen the exclusion power (compared to measurements
  based on solely pion sources). Their significance is particularly
  important for $\cos\delta\sim0$.
        \end{itemize}

A more specific question that we posed is whether, in case that the CP
violating phase $\delta$ is large ($\sim\pi/2$), the measurements of
flavor ratios among neutrino fluxes from astrophysical sources can
establish that the phase is indeed different from $0$ or $\pi$, and by
that prove that CP is violated in neutrino interactions.
Our conclusions regarding this question are the following:
\begin{itemize}
\item $\sin\theta_{13}$ must be large, between current $1-2\sigma$
  upper bounds.
\item The neutrino flux must not be lower than the
  Waxman-Bahcall bound. If the flux is smaller, a larger neutrino
  telescope may still achieve this goal, within a reasonable time scale ($\lesssim10$ years).
\item Neutrino flux from muon-damped sources must be identified, and
  the related flavor ratios measured with accuracy better than 10\%.
\item The uncertainties on $\theta_{12}$ and $\theta_{23}$ must be
  reduced by other experiments by a factor of about two.
        \end{itemize}
Even if all these conditions are met, the probability of excluding CP
conservation in neutrino oscillations is at best 60\%.

The strongest sensitivity to $\cos\delta$ arises in flavor ratios
related to muon-damped sources. On the theoretical side, a more
careful study of the transition at high energy from pion-source to
muon-damped source is important for better understanding of this
crucial ingredient in our analysis \cite{Lipari:2007su}. On Nature's
side, the lower the transition energy, and the sharper the transition,
the higher statistics of events from muon-damped source that will
become available and, consequently, the better chances are that a
neutrino telescope will contribute significantly to understanding CP
violation in neutrino oscillations. Finally, on the experimental side,
a neutrino telescope that is effectively ten times bigger than
IceCube, for neutrino energy $\sim100\ TeV$ (see Section
\ref{sec:neg}), is well motivated by our arguments.

The fact that establishing CP violation in IceCube, an experiment
under construction, is not manifestly impossible is exciting. While a
combination of several favorable circumstances is required to achieve
such a goal, it is worth to refine this analysis, to prepare for a
fortunate case that these circumstances are fulfilled by the
parameters of Nature and by the capabilities of neutrino telescopes.

%%%%%%%%%%%%%%%%%%%%%%%%%%%%%
\section*{Acknowledgments}
We are grateful to Concha Gonzalez-Garcia, Francis Halzen, and Walter
Winter for useful discussions. The research of Y.N. is supported by
the Israel Science Foundation founded by the Israel Academy of
Sciences and Humanities, the United States-Israel Binational Science
Foundation (BSF), Jerusalem, Israel, the German-Israeli foundation for
scientific research and development (GIF), and the Minerva Foundation.
E.W.'s research is partly supported by the ISF, AEC and Minerva
grants.

%%%%%%%%%%%%%%%%%%%%%%%%%%%%%

%%%%%%%%%%%%%%%%%%


\end{document}